\documentclass[11pt]{article}
\usepackage[margin=1in]{geometry}                % See geometry.pdf to learn the layout options. There are lots.
\usepackage[parfill]{parskip}    % Activate to begin paragraphs with an empty line rather than an indent
\usepackage{amssymb}
\usepackage{mathtools}

\usepackage{indentfirst}
\usepackage{verbatim}
\usepackage{amsmath}
\usepackage{amsthm}
\usepackage{amssymb}
\usepackage{bm}
\usepackage{dsfont}
\usepackage{lscape}
\usepackage{graphicx}
\usepackage{wrapfig}
\usepackage{verbatim}
\usepackage{helvet}
\usepackage{float}
\usepackage{calc}
\usepackage{xspace}
\usepackage{pifont}
\usepackage{caption}
\usepackage[caption=false]{subfig}
\usepackage{multirow}
\usepackage{hyperref}
\usepackage{placeins}
\usepackage[normalem]{ulem}
\usepackage{epstopdf}
\usepackage{movie15}
\usepackage{topcapt}
\usepackage[boxed]{algorithm2e}
\usepackage{hhline}

\usepackage{rotating}

\def\proof{\noindent{\bf Proof:} }
\def\qed{\hfill{$\Box$} \\}

\def\real{{\mathbb R}}
\def\nat{{\mathbb N}}

\def\hilbert{{\mathcal H}}

\def\herm{{\mathsf{H}}}
\def\trans{{\mathsf{T}}}

\usepackage{catchfile}
\usepackage{subdepth}

\newcommand{\getvar}[2]{%
  \CatchFileEdef#1{"|kpsewhich -var-value #2"}{\endlinechar=-1 }%
}

\def\me{yiyin}
\getvar{\usrtest}{USER}

\ifx\me\usrtest

\else

\fi

\newtheorem{theorem}{Theorem}

\newtheorem{lemma}{Lemma}

\newtheorem{definition}{Definition}
\newtheorem{remark}{Remark}
\newtheorem{example}{Example}
\newtheorem{algorithmp}{Algorithm}

\usepackage{upgreek} 
\usepackage{paralist}
\usepackage{xcolor}

\definecolor{h2blue}{RGB}{188,188,255}
\definecolor{h1green}{RGB}{188,255,188}

\usepackage{authblk}

\title{Sparse Identification of Contrast Gain Control in the Fruit Fly Photoreceptor and Amacrine Cell Layer}
\author[]{Aurel~A.~Lazar}
\author[]{Nikul~H.~Ukani}
\author[]{Yiyin~Zhou\thanks{The author's names are alphabetically listed.\newline Correspondence: Aurel A. Lazar, \newline Department of Electrical Engineering, \newline Columbia University,
\newline 500 West 120th Street, \newline New York, NY 10027, United States \newline Email: aurel@ee.columbia.edu (AAL), nikul@ee.columbia.edu (NHU), yiyin@ee.columbia.edu (YZ)}}
\affil[]{Department of Electrical Engineering, \\ Columbia University, New York, NY 10027}
\begin{document}
\maketitle

\begin{abstract}
\noindent
The fruit fly's natural visual environment is often characterized by light 
intensities ranging across several orders
of magnitude and by rapidly varying contrast  across space and time.
Fruit fly photoreceptors robustly transduce and, in conjunction with amacrine cells,  process visual scenes and provide the
resulting signal to downstream targets.
Here we model the first step of visual
processing in the photoreceptor-amacrine cell layer. 
We propose a novel divisive normalization processor (DNP)  for modeling the computation taking place in the photoreceptor-amacrine cell layer.
The DNP explicitly models the photoreceptor feedforward and temporal feedback processing paths 
and the spatio-temporal feedback path of the amacrine cells.
We then formally characterize the contrast gain control of the DNP
and provide sparse identification algorithms 
that can efficiently identify each the feedforward and feedback DNP components.
The algorithms presented here are the first demonstration of tractable and robust identification of the components of a divisive normalization processor. 
The sparse identification algorithms  
can be readily employed in experimental settings,
and their effectiveness is demonstrated with several examples.

\end{abstract}

\newpage
\tableofcontents
\newpage

\section{Introduction}
Sensory processing systems in the brain extract relevant information from inputs  whose amplitude can vary orders of magnitude \cite{LAUGHLIN1994, Rodieck1998, deBruyne4520, OLSEN2010}.
Consequently, at each layer of processing, starting right from sensory transduction, neurons need to map their output into a range that can be effectively processed by subsequent neural circuits.
As an example, photoreceptors \cite{SPB12, Meister1999, Scholl2012, BM2002} and olfactory receptor neurons \cite{Wilson2013, Firestein2001} in both vertebrates and invertebrates, adapt to a large range of intensity/temporal contrast values of visual and odorant stimuli. 
Adaptation to mean and variance of the stimuli has been observed in the  auditory system \cite{Clemens2018fe, Rabinowitz2011} as well. Further down the visual pathway, motion sensitive neurons in vertebrates and invertebrates, have been shown to be robust at various brightness and contrast levels \cite{Kohn2003, Rien2012, Heuer2002}.

Early visual circuits such as the photoreceptor/amacrine cell layer of the fruit fly brain are believed to perform spatio-temporal intensity and contrast gain control for dynamic adaptation to visual stimuli whose intensity and contrast vary orders of magnitude both in space and time. However, current theoretical methods of describing spatio-temporal gain control lack a systematic framework for characterizing its dynamics and identification algorithms to estimate circuit components are currently not available.

Divisive normalization \cite{CH2012} has been proposed as a canonical circuit model of computation for many sensory processing circuits underlying adaptation and attention. However, there is a lack of general mathematical framework for identifying such computations from recorded data.
In this paper we model the phoreceptor/amacrine cells layer of the fruit fly as  a feedforward and feedback temporal and spatio-temporal divisive normalization processor. We provide efficient algorithms for identifying all the components of the temporal as well spatio-temporal divisive normalization  processors.

This manuscript is organized as follows.
In Section~\ref{sec:dnp} the overall architecture of the divisive normalization processor (DNP) is introduced and its power of modeling contrast gain control
demonstrated. 
We first describe in Section~\ref{sec:bio_model} the biological model of photoreceptor-amacrine cell layer.
In Section~\ref{sec:div_norm_t}, 
 we introduce a general model for divisive normalization in the time domain.
The temporal DNP  consists of the ratio of two non-linear functionals acting on the input stimulus.
In Section \ref{sec:STDNP} we then extend the model to space-time domain to include models of
lateral feedback from amacrine cells, and
demonstrate its processing power.
In Sections~\ref{sec:iden_norm} and \ref{sec:space-time}, we provide identification algorithms and show that the temporal and spatio-temporal DNPs can be efficiently identified. 
We demonstrate the effectiveness of the algorithms with several examples.
We conclude the paper with a discussion in
Section~\ref{sec:disc}.

\section{The Architecture of Divisive Normalization Processors}
\label{sec:dnp}

In section \ref{sec:bio_model} we start by motivating the present
work.
We then introduce the architecture of
divisive normalization processors in the time domain
(section \ref{sec:div_norm_t}) and space-time
domain (section \ref{sec:STDNP}).
Finally, in Appendix 
\ref{sec:I/O_lateral} we provide examples that
characterize the I/O mapping of the class of temporal and spatio-temporal divisive normalization processors previously described.

\subsection{Modeling the Photoreceptors and Amacrine Cells Layer}
\label{sec:bio_model}

In what follows, we anchor the model description around the photoreceptor-amacrine cell layer of the fruit fly. The fly retina consists of $\sim800$ ommatidia, each of which hosts photoreceptors whose axons terminate in a secondary neuropil called lamina. There, they provide inputs to columnar Large
Monopolar Cells (LMCs) that project to the third visual neuropil, and to amacrine cells \cite{FD1989}.
Amacrine cells are interneurons that innervate
axon terminals of multiple photoreceptors. The photoreceptors, in turn, receive lateral feedback
from the amacrine cells as well as feedback from LMCs such as L2 neurons \cite{NZW2009b, rivera-alba_wiring_2011, Simon2015}.

\begin{figure}[h]
	\centering
\includegraphics[width=0.6\textwidth]{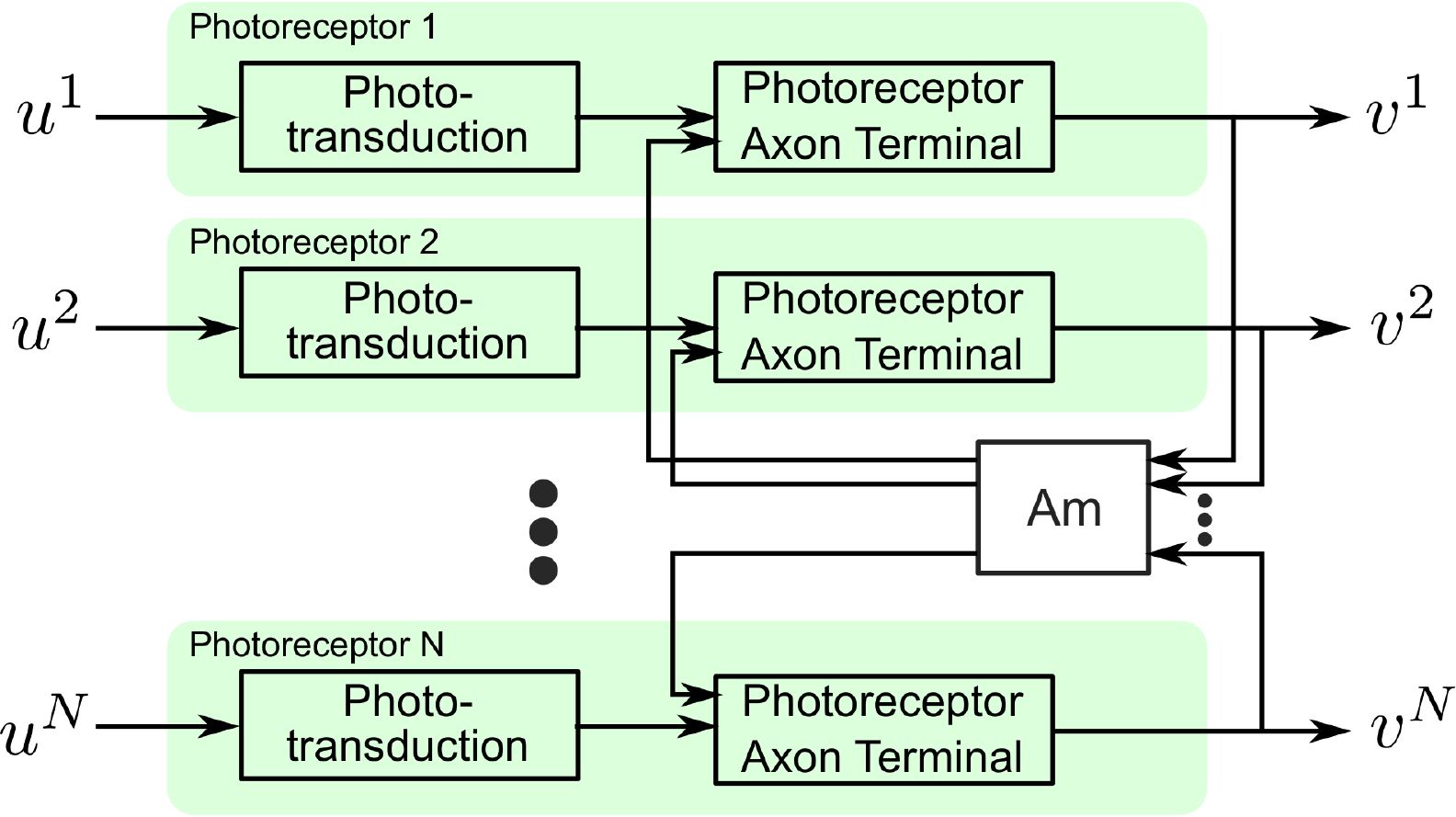}
	\caption{A schematic diagram of interaction between Amacrine cells and photoreceptors in multiple cartridges.}
	\label{fig:retina_circuit}
\end{figure}

A circuit diagram of the photoreceptor-amacrine cell layer is shown in
Figure~\ref{fig:retina_circuit}.
For the sake of clarity we assume here that an ommatidium consists of a single photoreceptor.
It has been shown that the outputs of photoreceptors exhibit rapid gain control through
both the phototransduction process and the interaction with such feedback loops \cite{ML81, JH2001, ZPW06, laughlin10, FBH16}.

In what follows we propose a model comprising non-linear transformations combined with divisive normalization that can model such gain control and can account for diverse dynamics.  We will also show that the model we propose can be systematically identified from observed input output pairs.

\subsection{Divisive Normalization Processors in the Time Domain}
\label{sec:div_norm_t}

In this section we present the modeling of temporal stimuli in \ref{sec:temp_stim_model}
and introduce a class of temporal divisive normalization processors for modeling photoreceptors in section \ref{sec:tdnp}.

\subsubsection{Modeling Temporal Stimuli}
\label{sec:temp_stim_model}

We model the temporal varying stimuli
$u_1=u_1(t)$, $t \in \mathbb{D} \subseteq \real$, to be real-valued
elements of the space of trigonometric polynomials \cite{LPZ2010}.
The choice of the space of the trigonometric polynomials has, as
we will see, substantial computational advantages.
A temporal stimulus models the visual field arising at the input of a single photoreceptor.
\begin{definition}
The space of trigonometric polynomials 
$\hilbert_1$ is the Hilbert space of
complex-valued functions
\begin{equation}
u_1(t) = \sum_{l=-L}^{L} a_{l} \cdot e_{l}(t),
\label{eq:uform}
\end{equation}
over the domain $\mathbb{D} =  [0, S]$, where
\begin{equation}
e_{l}(t) =  \frac{1}{\sqrt{S}}\operatorname{exp}\left(\frac{jl\Omega}{L}t \right).
\end{equation}
Here $\Omega$ denotes the bandwidth,
and $L$ the order of the space.
Stimuli $u_1\in\hilbert_1$ are extended to be periodic over $\real$ with period
$S = \frac{2\pi L}{\Omega}$.
\end{definition}

$\hilbert_1$ is a Reproducing Kernel Hilbert Space (RKHS)
\cite{BTA04} with
reproducing kernel (RK)
\vspace{-0.075in}
\begin{equation}
K_1(t;t') = \sum_{l=-L}^{L}e_{l}(t-t').
\end{equation}

We denote the dimension of $\hilbert_1$ by $dim(\hilbert_1)$ and 
$dim(\hilbert_1) = 2L+1$.

\begin{definition}
The tensor product space $\hilbert_2=\hilbert_1 \otimes \hilbert_1$ is an RKHS with reproducing kernel
\vspace{-0.1in}
\begin{equation}
K_2(t_1,t_2;t'_1,t'_2) = \sum_{l_{1}=-L}^{L}\sum_{l_{2}=-L}^{L}e_{l_{1}}(t_1-t'_1) \cdot e_{l_{2}}(t_2-t'_2).
\end{equation}
\end{definition}

Note that $dim(\hilbert_2) = dim(\hilbert_1)^2 =
(2 L +1)^2$.

\subsubsection{Temporal Divisive Normalization Processors}
\label{sec:tdnp}

\begin{figure}[htbp]
\centering
\includegraphics[width=0.6\linewidth]{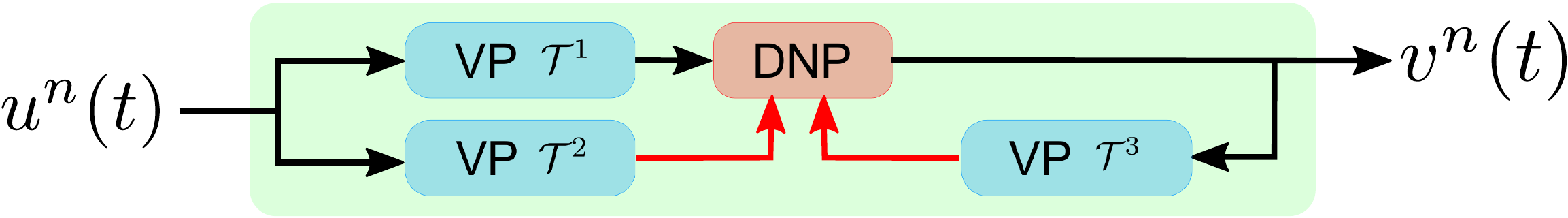}
\caption{Schematic diagram of a temporal divisive normalization processor.}
\label{fig:div_norm_t}
\end{figure}

We first consider single photoreceptors without feedback from
the amacrine cells shown in Figure \ref{fig:retina_circuit}.
A schematic of the temporal divisive normalization processor (DNP) modeling the photoreceptor  is shown in Figure \ref{fig:div_norm_t}.
For notational simplicity, we consider a single photoreceptor here.
The input visual stimulus to the photoreceptor is denoted
by $u=u(t), t\in \mathbb{D}$, and the output
electric current by $v=v(t), t\in \mathbb{D}$.

\begin{remark}
Note that in a single photoreceptor, photons are first absorbed by a large number of microvilli \cite{JH2001} (not shown). Microvilli generate  ``quantum bumps" in response to 
photons; the photoreceptor aggregates the bumps and in the process creates 
the transduction current. Calcium ion influx and calcium diffusion into the photoreceptor cell body may change
the sensitivity of the transduction cascade.
A high concentration of calcium (buffer) can result in a photon to be ineffective  and may also affect the duration and the magnitude of quantum bumps \cite{PGR08}.
\end{remark}

The DNP consists of 1) a feedforward Volterra Processor (VP) $\mathcal{T}^1$,
2) a feedforward normalization VP $\mathcal{T}^2$,
and 3) a feedback normalization VP $\mathcal{T}^3$.
The output of the photoreceptor amounts
to 
\begin{equation}
%v^i = \frac{\mathcal{T}^i_1 u}{\mathcal{T}^i_2 u + \mathcal{T}^i_3 v},
v^n = \frac{\mathcal{T}^1 u^n}{\mathcal{T}^2 u^n + \mathcal{T}^3 v^n}, \; n=1,2,\cdots,N,
\label{eq:io_n1}
\end{equation}
where
\begin{equation}
%(\mathcal{T}^i_l u)(t) = b^i_l + \int_{\mathbb{D}} h^{i}_{1l}(s) u(t-s) ds + \int_{\mathbb{D}^2} h^{i}_{2l}(s_1,s_2)u(t-s_1)u(t-s_2)ds_1ds_2, l = 1,2,
(\mathcal{T}^l u^n)(t) = b^l + \int_{\mathbb{D}} h_{1}^{l}(s) u^n_1(t-s) ds + \int_{\mathbb{D}^2} h_{2}^{l}(s_1,s_2)u^n_1(t-s_1)u^n_1(t-s_2)ds_1ds_2, l = 1,2,
\label{eq:vp12}
\end{equation}
and
\begin{equation}
%(\mathcal{T}^i_3 v)(t) = b^i_3 + \int_{\mathbb{D}} h^{i}_{13}(s) v(t-s) ds + \int_{\mathbb{D}^2} h^{i}_{23}(s_1,s_2)v(t-s_1)v(t-s_2)ds_1ds_2 .
(\mathcal{T}^3 v^n)(t) = b^3 + \int_{\mathbb{D}} h_{1}^{3} (s) v^n_1(t-s) ds + \int_{\mathbb{D}^2} h_{2}^{3} (s_1,s_2) v^n_1(t-s_1)v^n_1(t-s_2)ds_1ds_2 .
\label{eq:vp3}
\end{equation}
Here $b^{l}, l=1,2,3$, are the zero${}^{th}$-order
Volterra kernels (constants), $h_{1}^{l}(t), l=1,2,3$, are first-order Volterra kernels (impulse responses of linear filters), and 
$h_{2}^{l}(t,s), l=1,2,3$, second-order Volterra kernels.
As before, 
$\mathbb{D}$ denotes the domain of the input space, and $\mathbb{D}^2=\mathbb{D}\times\mathbb{D}$.
\begin{remark}
For the sake of tractability, we limit each nonlinear functional in \eqref{eq:vp12}  and \eqref{eq:vp3} to be 
composed of only a first and second-order Volterra kernels. The division in \eqref{eq:io_n1}
allows us, however, to model nonlinear processing of much higher orders.
\end{remark}

We note that $v$ in \eqref{eq:io_n1} is invariant under scaling by the same factor of the 
numerator and denominator.
Hence, without loss of generality, we will assume
$b^2+b^3=1$.
We will also assume that the DNP is bounded-input bounded-output \cite{Rugh1981}.

\subsubsection{Modeling Temporal DNP Feedback Filters}

Here we define the filter kernels in equations
\eqref{eq:vp12} and \eqref{eq:vp3}.

\begin{definition}
Let $h^l_p \in \mathbb{L}^1(\mathbb{D}^p), l=1,2, p=1,2$, where $\mathbb{L}^1$ denotes the space of Lebesgue integrable functions. The operator $\mathcal{P}_1: \mathbb{L}_1(\mathbb{D}) \rightarrow \hilbert_1$ given by
\begin{equation}
(\mathcal{P}_1 h^l_1)(t) = \int_{\mathbb{D}} h^l_1(t') K_1(t; t') dt'
\end{equation}
is called the projection operator from $\mathbb{L}^1(\mathbb{D})$ to $\hilbert_1$.
Similarly, the operator $\mathcal{P}_2: \mathbb{L}_1(\mathbb{D}^2) \rightarrow \hilbert_2$ given by
\begin{equation}
(\mathcal{P}_2h^l_2)(t_1;t_2) = \int_{\mathbb{D}^2} h^l_2(t'_1; t'_2) K_2(t_1,t_2; t'_1,t'_2) dt'_1dt'_2
\end{equation}
is called the projection operator from $\mathbb{L}^1(\mathbb{D}^2)$ to $\hilbert_2$.
\end{definition}
Note, that for $u_1^n \in \hilbert_1, \mathcal{P}_1 u_1^n = u_1^n$. Moreover, with $u_2^n(t_1,t_2) = u_1^n(t_1)u_1^n(t_2),  \mathcal{P}_2 u_2^n = u_2^n$.
Thus,
\begin{equation}
(\mathcal{T}^l u^n)(t) = b^l + \int_{\mathbb{D}} (\mathcal{P}_1 h_{1}^{l})(s) u^n_1(t-s) ds + \int_{\mathbb{D}^2} (\mathcal{P}_2 h_{2}^{l})(s_1,s_2)u^n_1(t-s_1)u^n_1(t-s_2)ds_1ds_2, l = 1,2,
\label{eq:vp12P}
\end{equation}
and by assuming that $h^l_1 \in \hilbert_1$, $l=1,2,$ and
$h^l_2 \in \hilbert_2$ we recover the simple form
of equation \eqref{eq:vp12}.

We model the output waveforms
$v_1^o = v_1^o (t)$, $t \in \mathbb{D} \subseteq \real$, to be real-valued
elements of the space of trigonometric polynomials \cite{LPZ2010}.
\begin{definition}
The space of trigonometric polynomials 
$\hilbert_1^o$ is the Hilbert space of
complex-valued functions
\begin{equation}
v_1^o(t) = \sum_{l=-L^o}^{L^o} a_{l}^o \cdot e_{l}^o (t),
\label{eq:uform}
\end{equation}
over the domain $\mathbb{D} =  [0, S]$, where
\begin{equation}
e_{l}^o (t) =  \frac{1}{\sqrt{S^o}}\operatorname{exp}\left(\frac{jl\Omega^o}{L^o}t \right).
\end{equation}
Here $\Omega^o$ denotes the bandwidth,
and $L^o$ the order of the space.
The output waveforms $v_1^o\in\hilbert_1^o$ are extended to be periodic over $\real$ with period
$S^o = \frac{2\pi L^o}{\Omega^o}$.
\end{definition}

$\hilbert_1^o$ is a Reproducing Kernel Hilbert Space (RKHS)
\cite{BTA04} with
reproducing kernel (RK)
\vspace{-0.075in}
\begin{equation}
K_1^o(t;t') = \sum_{l=-L^o}^{L^o} e_{l}^o(t-t').
\end{equation}

We denote the dimension of $\hilbert_1^o$ by $dim(\hilbert_1^o)$ and 
$dim(\hilbert_1^o) = 2L^o+1$.
\begin{definition}
The tensor product space $\hilbert_2^o = \hilbert_1^o \otimes \hilbert_1^o$ is an RKHS with reproducing kernel
\vspace{-0.1in}
\begin{equation}
K_2^o(t_1,t_2;t'_1,t'_2) = \sum_{l_{1}=-L^o}^{L^o}\sum_{l_{2}=-L^o}^{L^o}e_{l_{1}}^o(t_1-t'_1) \cdot e_{l_{2}}^o(t_2-t'_2).
\end{equation}
\end{definition}

Note that $dim(\hilbert_2^o) = dim(\hilbert_1^o)^2
= (2 L^o +1)^2$.

\begin{definition}
Let $h^3_p \in \mathbb{L}^1(\mathbb{D}^p), p=1,2$, where $\mathbb{L}^1$ denotes the space of Lebesgue integrable functions. The operator $\mathcal{P}_1^o: \mathbb{L}_1(\mathbb{D}) \rightarrow \hilbert_1^o$ given by
\begin{equation}
(\mathcal{P}_1^o h^3_1)(t) = \int_{\mathbb{D}} h^3_1(t') K_1^o(t; t') dt'
\end{equation}
is called the projection operator from $\mathbb{L}^1(\mathbb{D})$ to $\hilbert_1$.
Similarly, the operator $\mathcal{P}_2^o: \mathbb{L}_1(\mathbb{D}^2) \rightarrow \hilbert_2$ given by
\begin{equation}
(\mathcal{P}_2^o h^3_2)(t_1;t_2) = \int_{\mathbb{D}^2} h^3_2(t'_1; t'_2) K_2^o(t_1,t_2; t'_1,t'_2) dt'_1dt'_2
\end{equation}
is called the projection operator from $\mathbb{L}^1(\mathbb{D}^2)$ to $\hilbert_2^o$.
\end{definition}

We note that 
\begin{equation}
(\mathcal{T}^3 v^n)(t) = b^3 + \int_{\mathbb{D}} (\mathcal{P}_1^o h_{1}^{3}) (s) v^n_1(t-s) ds + \int_{\mathbb{D}^2} (\mathcal{P}_2^o h_{2}^{3}) (s_1,s_2) v^n_1(t-s_1)v^n_1(t-s_2)ds_1ds_2 .
\label{eq:vp3P}
\end{equation}
If we now assume that $h^3_1\in \hilbert^o_1$ and $h^3_2 \in \hilbert^o_2$, and thereby $\mathcal{P}_1^o h_1^3 = h_1^3$ and $\mathcal{P}_2^o h_2^3 = h_2^3$, respectively,
and the above equation is identical with equation
\eqref{eq:vp3} above.

\subsection{Divisive Normalization Processors in the Space-Time Domain}
\label{sec:STDNP}

In Section~\ref{sec:div_norm_t}, we described a temporal divisive normalization processor model.
The normalization term was the sum of a processed version of the input and the output.
However, many biological circuits are thought to exhibit lateral inhibition and gain control 
\cite{VanLeeuwen2009, BCG1970, Olsen2008, PS1993, CH2012}.
An example is provided by the photoreceptor-amacrine cell layer shown in Figure \ref{fig:retina_circuit}. 
In Figure \ref{fig:div_norm_st}, we provide
a model of the schematic in Figure
\ref{fig:retina_circuit}. This model is 
a `circuit' extension of the one shown in Figure \ref{fig:div_norm_t} and accounts explicitly for lateral inhibition.
We anchor the extension in the visual system where the
input domain is spatio-temporal.

In section \ref{sec:msts} below
we model spatio-temporal stimuli and in section \ref{sec:stdnp} the spatio-temporal divisive normalization processors.

\subsubsection{Modeling Spatio-Temporal Stimuli}
\label{sec:msts}

In this section we provide a model of the interaction between a group of photoreceptors and an amacrine cell.
In each photoreceptor, the phototransduction process converts light
into current and excites the membrane of the photoreceptor. The voltage
signal is then propagated through its axon to the lamina. While photoreceptors
provide inputs to amacrine cells, their axon terminals also receive amacrine cell
input. Since an amacrine cell innervates multiple lamina cartridges, it
provides spatial feedback to several photoreceptors
in a small neighborhood.

We extend the temporal divisive normalization processor depicted in Figure~\ref{fig:div_norm_t} to process
spatio-temporal stimuli as shown in Figure~\ref{fig:div_norm_st}. 
Each spatially sampled point, or pixel, denoted as $u^i(t), i=1,2,\cdots,N$, is first processed by a temporal DNP.
For simplicity rather than picking different
Volterra kernels for each branch, $\mathcal{T}^1, \mathcal{T}^2$ and $\mathcal{T}^3$ are shared across branches.
In addition, we introduce a Multi-Input Volterra Processor (MVP) to model
the spatio-temporal  feedback due to the amacrine cell.
Each of the branches in Figure~\ref{fig:div_norm_st}, without the input of the MVP block, is equivalent 
with the model in Figure~\ref{fig:div_norm_t}.

\subsubsection{Spatio-Temporal Divisive Normalization Processors}
\label{sec:stdnp}

\begin{figure}[t!]
\centering
%\subfloat[]{
\includegraphics[width=0.7\linewidth]{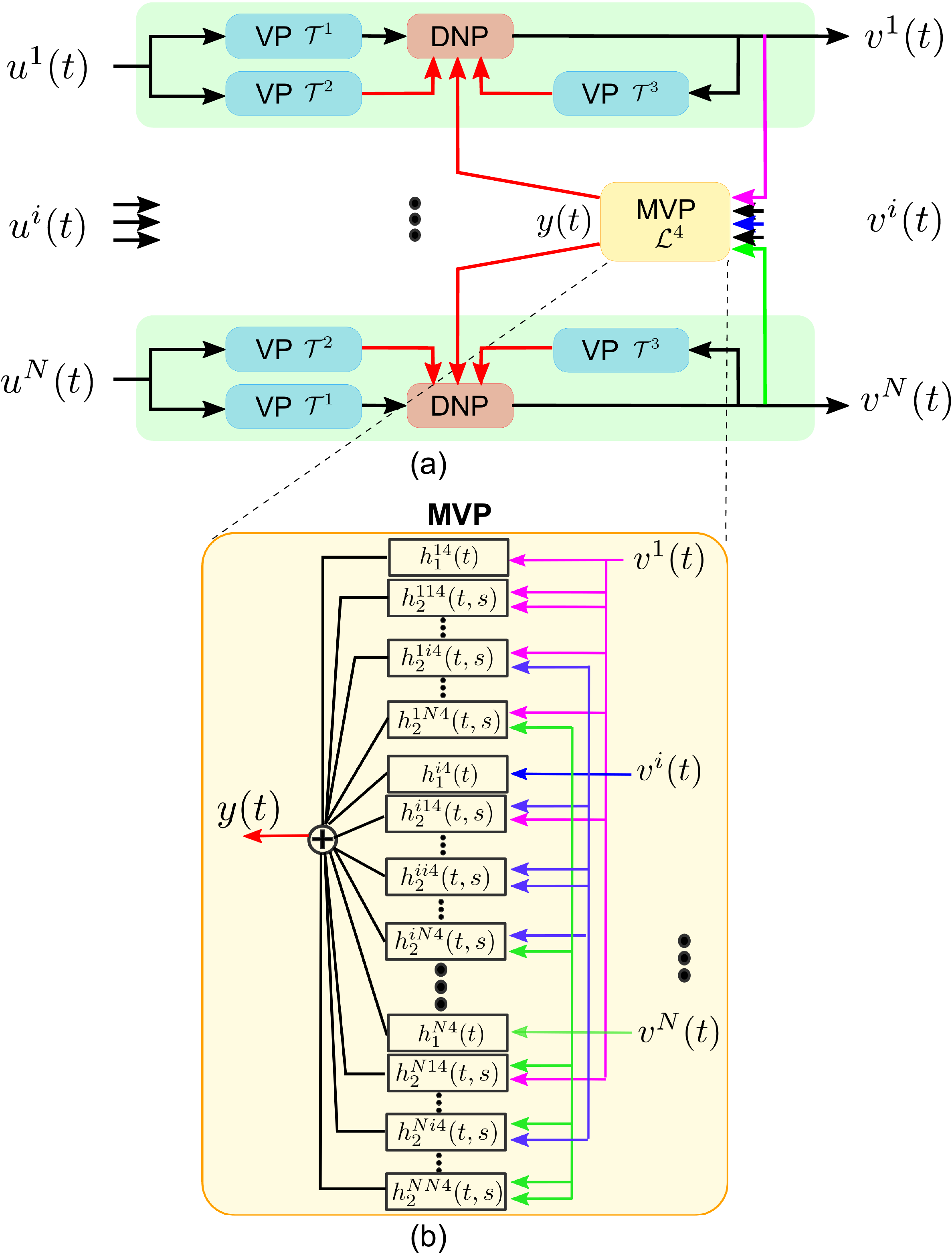}
%\subfloat[]{\label{fig:mvp}
%\includegraphics[width=0.3\linewidth]{figures/mvp.pdf}}
\caption{Schematic block diagram of the spatio-temporal divisive normalization processor.}
\label{fig:div_norm_st}
%\label{fig:divnorm_spatio_temporal}
\end{figure}

As shown in Figure~\ref{fig:div_norm_st}(b), MVPs are comprised of second-order filters acting on DNP output pairs in 
addition to linear filters independently  acting on each DNP output.
Thus, the inputs to the MVP are the DNP outputs  $v^i(t), i = 1,2,\cdots,N$,
and the MVP output amounts to
\begin{equation}
\begin{split}
%\left(\mathcal{L}_4\mathbf{v}\right)(t) = & b_4 +  \sum_{i=1}^{N}\sum_{j=1}^{M} \left(\int_{\mathbb{D}}h^{ij}_1(s)v(x_i, y_j, t-s)ds \right) \\
%&+ \sum_{i=1}^{N}\sum_{j=1}^M \sum_{k=1}^{N}\sum_{l=1}^M  \left(\int_{\mathbb{D}^2} h_2^{ijikl}(t-s_1,t-s_2)v(x_i, y_j, t-s_1)v(x_{k}, y_{l}, t-s_2)ds_1ds_2\right) ,
\left(\mathcal{L}^4\mathbf{v}\right)(t) = %&
b^4 +  \sum_{i=1}^{N}\left(\int_{\mathbb{D}} h^{i4}_{1} (s)v^i(t-s)ds \right)
%\\ &
+ \sum_{i=1}^{N}\sum_{j=1}^{N}  \left(\int_{\mathbb{D}^2} h_{2}^{ij4} (s_1,s_2)v^i(t-s_1)v^j(t-s_2)ds_1ds_2\right) ,
\end{split}
\end{equation}
where
\begin{equation}
%\mathbf{v}(t) = \left[v(x_0, y_0,t), v(x_1, y_0, t), \cdots, v(x_N, y_0, t), \cdots, \cdots, v(x_0, y_M, t), v(x_1, y_M, t), \cdots, v(x_N, y_M, t)\right]^T.
\mathbf{v}(t) = \left[v^1(t), v^2(t), \cdots, v^N(t)\right]^T,
\end{equation}
$b^4$ is the zeros${}^{th}$-order Volterra kernel (constant).
Furthermore, $h^{i4}_1\in\hilbert^o_1, i=1,2,\cdots,N$, are the first-order Volterra kernels whose inputs are $v^i, i=1,2,\cdots,N$, respectively, and
$h^{ij4}_2\in\hilbert^o_2, i,j=1,2,\cdots,N$, are the second-order Volterra kernels whose
inputs are the pairs $(v^i, v^j), i,j=1,2,\cdots,N$, respectively.

The full model in Figure \ref{fig:div_norm_st} thus consists of parallel channels as depicted in 
Figure~\ref{fig:div_norm_t} with the added cross-channel 
feedback normalization/gain control provided by the MVP block.

The overall output of the spatio-temporal DNP can be expressed as 
\begin{equation}
%v(x_i, y_j, t) = \frac{\mathcal{T}_1 u(x_i, y_j, t)}{\mathcal{T}_2 u(x_i, y_j,t) + \mathcal{T}_3 v(x_i, y_j, t) + \mathcal{L}_4 \mathbf{v}}, \; i=1,2,\cdots,N, \; j = 1,2,\cdots,M.
v^n(t) = \frac{\mathcal{T}^1 u^n}{\mathcal{T}^2 u^n + \mathcal{T}^3 v^n + \mathcal{L}^4 \mathbf{v}}, \; n=1,2,\cdots,N.
\label{eq:stdnp}
\end{equation}
W.L.O.G., we assume that $b^2 + b^3 + b^4 = 1$.

\subsubsection{Spatio-Temporal DNPs and Contrast Gain Control}

The relationship and some intuition behind the modeling power of spatio-temporal DNPs is provided in several examples Appendix~\ref{sec:I/O_lateral}. 
The I/O of 3 simple spatio-temporal DNPs stimulated with different inputs is briefly mentioned here.
In the first example, we evaluated the response of a DNP 
with 4 photoreceptors under different background light intensity levels.
In Figure~\ref{fig:io_flashes} one of the photoreceptors is subject to an additional flash of different light intensity, while the inputs of the other 3
are kept at the same background level.
The steady state response of the photoreceptor that receives the additional flash is shifted as a function of the background-intensity level.
In the second example, contrast gain control exerted by the amacrine cells is demonstrated for the same DNP in Figure~\ref{fig:lateral_bar_io}.
The effect of the MVP block on the RMS contrast can be clearly seen in Figure~~\ref{fig:lateral_bar_io} and is quantitatively evaluated in
Figure~\ref{fig:contrast_contrast}.
Finally, an example of steady state I/O visualization of  a natural image with resolution of $1,536\times 1,024$ pixels at low, medium and high luminance
values is shown in Figure~\ref{fig:lateral_natural_io}. 
The image was divided into $16\times 16$ spatio-temporal DNP blocks with a 4 pixel overlap in each direction.

\section{Sparse Identification of Temporal DNPs}
\label{sec:iden_norm}

In what follows we derive sparse identification algorithms for the components of spatio-temporal DNPs depicted in Figure~\ref{fig:div_norm_t} and formally defined in equation \eqref{eq:io_n1}.
In what follows, 
we assume that during experimental trials,  the single isolated photoreceptor $n$ is presented with $M$ test stimuli $u^{nm}_1 = u^{nm}_1(t), \;m=1, 2, \cdots,M$,  and for each trial the outputs $v^{nm}_1 = v^{nm}_1(t), m=1, 2, \cdots, M$, are recorded. The objective is to identify the model components $b_1, h^l_{1}, h^l_{2}, l = 1,2,3$, from the knowledge of the inputs and outputs.

\subsection{Deriving the Sparse Identification Algorithm for Temporal DNPs}

\begin{lemma}
With $M$ input stimuli presented to a temporal DNP,
let the inputs $u^{nm}(t)$ and the outputs $v^{nm}(t)$ be sampled at times
$(t_k), k = 1,2,\cdots,T$. Then,
\begin{equation}
b^1 + \left\langle h_{1}^{1}, \phi_{11}^{nmk} \right\rangle_{\hilbert_1} \! \! \! +
	 \left\langle h_{1}^{2} , \phi_{12}^{nmk} \right\rangle_{\hilbert_1} \! \! \! + 
	 \left\langle h_{1}^{3} , \phi_{13}^{nmk} \right\rangle_{\hilbert_1^o} \!  +
	 %\nonumber
	 %\\ &+
	 \left\langle h_{2}^{1}, 
	 \phi_{21}^{nmk} \right\rangle_{\hilbert_2}
	 \! \! \! + 
	 \left\langle h_{2}^{2}, \phi_{22}^{nmk}  \right\rangle_{\hilbert_2} \! \! \! +
	 \left\langle h_{2}^{3}, \phi_{23}^{nmk}  \right\rangle_{\hilbert_2^o}
	 = q^{nmk} ,
	 \label{eq:norm_inner_prod_form}
\end{equation}
where the sampling functions $\phi_{11}^{nmk} \in \hilbert_1, \;\; \phi_{12}^{nmk}\in\mathcal{H}_1,
 \;\; \phi_{13}^{nmk}\in\mathcal{H}_1^o, \;\; \phi_{21}^{nmk}\in\mathcal{H}_2, \;\; \phi_{22}^{nmk}\in\mathcal{H}_2 \mbox{ and } \phi_{23}^{nmk}\in\mathcal{H}_2^o$ are given by
\begin{align}
\phi_{11}^{nmk} (t) &= u^{nm}_1(t_k-t) , \nonumber\\
\phi_{12}^{nmk} (t)  &=  -q^{nm}(t_k) u^{nm}_1(t_k-t)  , \nonumber\\
\phi_{13}^{nmk} (t)  &=  -q^{nm}(t_k) (\mathcal{P}_1^o v^{nm}_1) (t_k-t)  , 
\label{eq:paxes} \\
\phi_{21}^{nmk} (t, s)  &=  u^{nm}_1(t_k-t)u^{nm}_1(t_k-s) ,  \nonumber\\
\phi_{22}^{nmk} (t, s)  &= -q^{nm}(t_k) u^{nm}_1(t_k-t)u^{nm}_1(t_k-s) , \nonumber\\
\phi_{23}^{nmk} (t, s)  &= -q^{nm}(t_k) (\mathcal{P}_1^o v^{nm}_1) (t_k-t)(\mathcal{P}_1^o v^{nm}_1)(t_k-s) ,
\nonumber
\end{align} 
and $q^{nmk} = v^{nm}(t_k)$ for all $m=1,2,...,M$ and $k=1,2,...,T$.
\label{thm:div_iden}
\end{lemma}
\proof See Appendix~\ref{sec:app_th1}.

\begin{remark}
	From \eqref{eq:norm_inner_prod_form}, it can be seen that the identification of the divisive normalization processor has been reformulated as a generalized sampling problem \cite{Christensen2008} in
$\real\oplus\hilbert_1\oplus\hilbert_1\oplus\hilbert_1^o\oplus\hilbert_2\oplus\hilbert_2\oplus\hilbert_2^o$. Subsequently, the divisive normalization model can be identified by solving a system of linear equations.
\end{remark}

In order to solve the system of linear equations in \eqref{eq:norm_inner_prod_form}
we rewrite them first in matrix form.
\begin{lemma}
Eq. \eqref{eq:norm_inner_prod_form} can be expressed in matrix form as
\begin{equation}
%\langle \mathbf{c}_1,  \boldsymbol\Phi^{mk} \rangle + 
%\langle \mathbf{C}_2, \boldsymbol\Xi^{mk} \rangle = 
 \mathbf{c}_1^\trans \boldsymbol\Phi^{nmk} 
+ \operatorname{Tr}\left(\mathbf{C}_2^\herm \boldsymbol\Xi^{nmk}\right) = q^{nmk} ,
\end{equation}
for all $m=1,2,\cdots,M$, and $k=1,2,\cdots,T$,
where $\operatorname{Tr}(\cdot)$ denotes the trace operator, and
\begin{align}
%    \begin{array}{ccc}
\mbox{Measurements} & \leftarrow  q^{nmk} \quad (scalar) \\
\mbox{Unknowns} & \leftarrow  \left\{
\begin{array}{c}
\mathbf{c}_1 = \begin{bmatrix} b^1 & (\mathbf{h}^1_1)^\trans & (\mathbf{h}^2_1)^\trans  & (\mathbf{h}^3_1)^\trans \end{bmatrix}^\trans_{(4L+2L^o+4)\times 1}\\
\mathbf{C}_2 = \begin{bmatrix}
\mathbf{H}_2^1& \mathbf{0}_{(2L+1)\times(2L^o+1)}\\
\mathbf{H}_2^2 & \mathbf{0}_{(2L+1)\times(2L^o+1)} \\
\mathbf{0}_{(2L^o+1)\times(2L+1)} & \mathbf{H}_2^3
 \end{bmatrix}_{(4L+2L^o+3) \times (2L+2L^o+2)},
\end{array}
\right. \label{eq:c1_c2T}\\
\mbox{Sampling vectors} & \leftarrow  \left\{
\begin{array}{c}
\boldsymbol\Phi^{nmk} = \begin{bmatrix} 1\\ \mathbf{u}^{nmk} \\ -q^{nmk}\mathbf{u}^{nmk} \\ -q^{nmk} \mathbf{v}^{nmk} \end{bmatrix}_{(4L+2L^o+4) \times 1} \\
\boldsymbol\Xi^{nmk} = 
\begin{bmatrix}
\mathbf{U}_2^{nmk} & \mathbf{0}_{(2L+1)\times(2L^o+1)} \\
-q^{nmk}\mathbf{U}^{nmk}_2 & \mathbf{0}_{(2L+1)\times(2L^o+1)} \\
\mathbf{0}_{(2L^o+1)\times(2L+1)} & -q^{nmk} \mathbf{V}_2^{nmk}
\end{bmatrix}_{(4L+2L^o+3) \times (2L+2L^o+2)}
\end{array}
\right.
%    \end{array}
\end{align}
\label{lem:matrix_form}
\end{lemma}

\proof See Appendix~\ref{sec:lem1} for more  notation and detailed proof.

A necessary condition on the number of trials and the number of measurements required for identifying the divisive normalization processor for solving the system of equations in Theorem~\ref{thm:div_iden} is that the number
of trials $M \ge 3 + 2 \cdot dim(\hilbert_1)$ and the number of total samples $TM \ge 1 + 2 \cdot dim(\hilbert_1) + 2 \cdot dim(\hilbert_1)^2$.

It is easy to see that solving the system of equations above suffers from the curse of dimensionality.
As the dimension of $\hilbert_1$ increases, the number of samples needed
to identify components increases quadratically. 
Note that
the second-order Volterra kernels $h_{2}^{l}, l = 1,2,3$, have unique symmetric forms with orthogonal expansions as \cite{Rugh1981}
\begin{equation}
%h^i_{2l}(t_1,t_2)	= \sum_{k=1}^{r^i_l} \lambda^{ik}_{l} g^{ik}_{1l}(t_1)g^{ik}_{1l}(t_2), \qquad\qquad \|g^{i}_l\| = 1.
h_{2}^{l}(t_1,t_2)	= \sum_{k\in\nat}  \lambda^{kl} g^{kl}_{1}(t_1)g^{kl}_{1}(t_2), \qquad\qquad \|g_1^{kl}\| = 1 ,
\label{eq:h2_gen_form}
\end{equation}
where $g_1^{kl} \in \hilbert_1, k \in \nat$, are orthogonal to each other.
In what follows, we assume that the second-order Volterra kernels are sparse, \textit{i.e.}, $\lambda^{kl}=0$ for $k>r_l$, where, $r_l \ll dim(\hilbert_1)$.
Sparse kernels often arise in modeling sensory processing, \textit{e.g.}, in complex cells 
in the primary visual cortex \cite{LUZ2018}. 
By exploiting the sparse structure of the second order kernels,
the identification problem can be made tractable.

The sparsity of the kernels can be translated into a low-rank condition on the matrix representation of $h^l_2, l=1,2,3$ (see also Appendix~\ref{sec:lem1}).
Ideally, the optimization problem would be a rank minimization problem. But rank minimization being NP-hard, we use the surrogate of nuclear norm minimization instead, which is the convex envelope of the rank operator \cite{FHB2004}. 

To perform sparse identification of the divisive normalization processor we devised  Algorithm 1 below.
By optimizing over $\mathbf{c}_1$ and $\mathbf{C}_2$ and subsequently assigning the corresponding block entries according to \eqref{eq:c1_c2T}, Algorithm 1 identifies $b_1$, $\mathbf{h}^i_1, i= 1,2,3$ and $\mathbf{H}^i_2,i =1,2,3$.

As a surrogate of rank minimization,  Algorithm~\ref{al:norm:iden} minimizes a linear combination of the 
nuclear norm of $\mathbf{C}_2$, and the Euclidean norm of $\mathbf{c}_1$.
The optimization constraints correspond
i) in \eqref{eq:constraint1} to the generalized sampling problem allowing certain amount of error,
ii) in \eqref{eq:constraint2} to zero mean slack variables,
iii) in \eqref{eq:constraint3} {to the zeros in the two blocks in the top-right of $\mathbf{C}_2$, 
iv) in \eqref{eq:constraint4} to the zeros in the block in the lower-left of $\mathbf{C}_2$, 
and v) in \eqref{eq:constraint5}, \eqref{eq:constraint6} and \eqref{eq:constraint7}, respectively, $\mathbf{H}^1_2$, $\mathbf{H}^2_2$ and $\mathbf{H}^3_2$ are Hermitian.

\begin{algorithmp}
$\widehat{\mathbf{c}}_1$ and $\widehat{\mathbf{C}}_2$ are the solution to the following optimization problem 
\begin{align}
& \underset{\mathbf{c}_1, \;\mathbf{C}_2,  \; \boldsymbol\varepsilon}{\text{minimize}}
& & \|\mathbf{C}_2\|_{*} + \lambda_1 \|\mathbf{c}_1\|_2
% + \lambda_1\operatorname{Tr}(\mathbf{C_2}) 
+ \lambda_2 \| \boldsymbol\varepsilon \|_2 \label{eq:minimize1}  \\
& \text{s.t.}
& & \mathbf{c}_1^\trans \boldsymbol\Phi^{nmk} 
+ \operatorname{Tr}\left(\mathbf{C}_2^\herm \boldsymbol\Xi^{nmk}\right) = q^{nmk} 
+ \varepsilon^{(m-1)*T+k}, m = 1,\cdots,M, k=1,\cdots,T, \label{eq:constraint1} \\
&&& \mathbf{1}^\trans{\boldsymbol\varepsilon} = {\bm 0} \label{eq:constraint2} \\
&&& \begin{bmatrix} 
\mathbf{I}_{2\dim(\mathcal{H}_1) } 
& \mathbf{0}
\end{bmatrix}
\mathbf{C_2}
\begin{bmatrix}
\mathbf{0} \\
\mathbf{I}_{\dim(\mathcal{H}_1^o)} 
\end{bmatrix}    =  \mathbf{0}, \label{eq:constraint3}  \\  %top-right 0
&&& \begin{bmatrix} 
 \mathbf{0}
& \mathbf{I}_{\dim(\mathcal{H}_1^0) } 
\end{bmatrix}
\mathbf{C_2}
\begin{bmatrix}
\mathbf{I}_{\dim(\mathcal{H}_1)} \\
\mathbf{0}
\end{bmatrix}    =  \mathbf{0},  \label{eq:constraint4} \\% bottom-right zero
&&&\begin{bmatrix} 
\mathbf{I}_{\dim(\mathcal{H}_1)}
& \mathbf{0}
\end{bmatrix}
\mathbf{C_2}
\begin{bmatrix}
\mathbf{I}_{\dim(\mathcal{H}_1)} \\
\mathbf{0} 
\end{bmatrix}    =
\begin{bmatrix}
\mathbf{I}_{\dim(\mathcal{H}_1)} \\
\mathbf{0}
\end{bmatrix}^\trans 
\mathbf{C_2}^\herm
\begin{bmatrix} 
\mathbf{I}_{\dim(\mathcal{H}_1)}
& \mathbf{0}
\end{bmatrix}^\trans \!\!,
\label{eq:constraint5} 
\\      %H^1 hermitian
&&& \begin{bmatrix} 
\mathbf{0}_{\dim(\mathcal{H}_1)} \!\!
& \mathbf{I}_{\dim(\mathcal{H}_1)} \!\!
& \mathbf{0}
\end{bmatrix}
\mathbf{C_2}
\begin{bmatrix}
\mathbf{I}_{\dim(\mathcal{H}_1)} \\
\mathbf{0}
\end{bmatrix}    = 
\begin{bmatrix}
\mathbf{I}_{\dim(\mathcal{H}_1)} \\
\mathbf{0}
\end{bmatrix}^\trans \!\!
\mathbf{C_2}^\herm
\begin{bmatrix} 
\mathbf{0}_{\dim(\mathcal{H}_1)} \!\!
& \mathbf{I}_{\dim(\mathcal{H}_1)}  \!\!
& \mathbf{0}
\end{bmatrix}^\trans \!\!,  \label{eq:constraint6} \\  %H^2 hermitian
&&& 
\begin{bmatrix} 
\mathbf{0}
& \mathbf{I}_{\dim(\mathcal{H}_1^o)}  
\end{bmatrix}
\mathbf{C_2}
\begin{bmatrix}
\mathbf{0} \\
\mathbf{I}_{\dim(\mathcal{H}_1^o)} 
\end{bmatrix}    =  
 \begin{bmatrix}
\mathbf{0} \\
\mathbf{I}_{\dim(\mathcal{H}_1^o)} 
\end{bmatrix}^\trans 
\mathbf{C_2}^\herm
\begin{bmatrix} 
\mathbf{0}
& \mathbf{I}_{\dim(\mathcal{H}_1^o)}  
\end{bmatrix}^\trans \!\!.
\label{eq:constraint7}
\end{align}
where $\|\cdot\|_*$ denotes the nuclear norm defined as $\|\mathbf{C}_2\|_* = \mbox{Tr}\left(\left(\mathbf{C}_2^\herm\mathbf{C}_2\right)^\frac{1}{2}\right)$, $\;\;\lambda_1, \lambda_2$ are appropriately chosen hyperparameters, $\varepsilon_{i}  \in \real$ represent slack variables, $\mathbf{1}$ represents a vector of all ones, $\mathbf{I}_{p}$ represents a $p\times p$ identity matrix, $\mathbf{0}_{p}$ represents a $p\times p$ matrix of all zeros, $\mathbf{0}_{p\times q}$ represents a $p\times q$ matrix of all zeros and $\mathbf{0}$ represents a matrix of all zeros with dimensions that makes the equation consistent.
\label{al:norm:iden}
\end{algorithmp}

\begin{theorem}
The filters of the DNP are identified as
$\hat{b}^1 = \hat{b}^1$ and
\begin{align}
&\widehat{h_1^1} (t) = \sum\limits_{l = -L}^{L} [\widehat{\mathbf{h}^1_1}]_{l+L+1} \cdot e_{l}(t),
 ~~~~~
\widehat{h_2^1} (t_1, t_2) = \sum\limits_{l_{1} = -L}^{L}\sum\limits_{l_{2} = -L}^{L} [\widehat{\mathbf{H}^1_2}]_{l_{1}+L+1,L+1-l_2} \cdot e_{l_{1}}(t_1) e_{l_{2}}(t_2),\\
&\widehat{h_1^2} (t) = \sum\limits_{l = -L}^{L} [\widehat{\mathbf{h}^2_1}]_{l+L+1} \cdot e_{l}(t),
 ~~~~~
 \widehat{h_2^2} (t_1, t_2) = \sum\limits_{l_{1} = -L}^{L}\sum\limits_{l_{2} = -L}^{L} [\widehat{\mathbf{H}^2_2}]_{l_{1}+L+1,L+1-l_{2}} \cdot e_{l_{1}}(t_1) e_{l_{2}}(t_2),\\\
&\widehat{h_1^3} (t) = \sum\limits_{l = -L^o}^{L^o} [\widehat{\mathbf{h}^3_1}]_{l+L+1} \cdot e^o_{l}(t),
~~~~
 \widehat{h_2^3} (t_1, t_2) = \! \sum\limits_{l_{1} = -L^o}^{L^o} \! \sum\limits_{l_{2} = -L^o}^{L^o} [\widehat{\mathbf{H}^3_2}]_{l_{1}+L^o+1,L^o+1-l_{2}} \cdot e^o_{l_{1}}(t_1) e^o_{l_{2}}(t_2),
 \end{align}
where
\begin{align}
\begin{bmatrix}
 				\hat{b}^1 &
 				\widehat{\mathbf{h}^1_1}^\trans &
 				\widehat{\mathbf{h}^2_1}^\trans &
				\widehat{\mathbf{h}^3_1}^\trans
 \end{bmatrix}^\trans = \widehat{\mathbf{c}}_1 
%\end{align}
~~~~~~~and~~~~~
%\begin{align}
\begin{bmatrix}
\widehat{\mathbf{H}_2^1}& - \\
\widehat{\mathbf{H}_2^2} & - \\
- & \widehat{\mathbf{H}_2^3} 
 \end{bmatrix} =  \widehat{\mathbf{C}}_2 .
 \label{eq:C2}
 \end{align}
\end{theorem}

\begin{remark}
By exploiting the structure of low-rank second-order Volterra kernels, Algorithm~\ref{al:norm:iden} provides a tractable solution to the identification of the components of the divisive normalization processor.
\end{remark}

\subsection{Examples of Sparse Identification of Temporal DNPs}

We provide here identification examples solved using Algorithm~\ref{al:norm:iden}.

\begin{example}
Here, we identify a temporal divisive normalization processor in Figure~\ref{fig:div_norm_t}, where
\begin{eqnarray*}
h^1_1(t) &=& 2.472\times 10^{10} t^3 e^{-100\pi t}cos(36\pi t) \\
h^2_1(t) &=& 3.117\times 10^{8} t^3 e^{-100\pi t} cos(20\pi t) \\
h^3_1(t) &=& 4.753\times 10^8 t^3 e^{-100\pi t} cos(52 \pi t) \\
h^1_2(t,s) &=& 9.038 \times 10^{19} t^3s^3 e^{-100\pi (t+s)} cos(52\pi t)cos(52\pi s) \\ &&+ 5.3467\times 10^{14} t^3s^3 e^{-100\pi (t+s)} cos(100\pi t)cos(100\pi s)  \\
h^2_2(t,s)  &=& 1.533 \times 10^{19} t^3s^3 e^{-100\pi (t+s)} cos(68\pi t)cos(68\pi s) \\
&&+ 5.970\times 10^{14} t^3s^3 e^{-100\pi (t+s)} cos(84\pi t)cos(84\pi s) \\
h^3_2(t,s)  &=& 6.771 \times 10^{19} t^3s^3 e^{-100\pi (t+s)} cos(100\pi t)cos(100\pi s) \\
&&+ 5.970\times 10^{16} t^3s^3 e^{-100\pi (t+s)} cos(84\pi t)cos(84\pi s) \\
\end{eqnarray*}
We choose the input space $\hilbert_1$ to have $L = 10, \Omega = 100\pi$. Thus $S = 0.2s$ ,  $dim(\hilbert_1) = 21$ and $dim(\hilbert_2) = 441$.
Note that all three second-order Volterra kernel exhibit low-rank structure.
We presented the model with 25 stimuli from $\hilbert_1$, whose coefficients were chosen to be i.i.d gaussian variables. Then, a total of $425$ measurements were used from the input and the observed output pairs to solve the identification problem using Algorithm~\ref{al:norm:iden}. The results of the identification are shown in Figure~\ref{fig:iden_norm}. As can be seen from the figure, Algorithm~\ref{al:norm:iden} was able to identify the model with high precision using only $450$ measurements, much less than the $1387$ measurements that would have been required to solve the generalized sampling problem directly. The factor of reduction in the required measurements is critical when the model needs to be identified in a much larger space, for example, a space of spatio-temporal stimuli as shown in the next example. 
\begin{figure}[htbp]
%	\hspace{-1in}
%\includegraphics[width=1.0\linewidth]{figures/identification_example.pdf}
\includegraphics[width=1.0\linewidth]{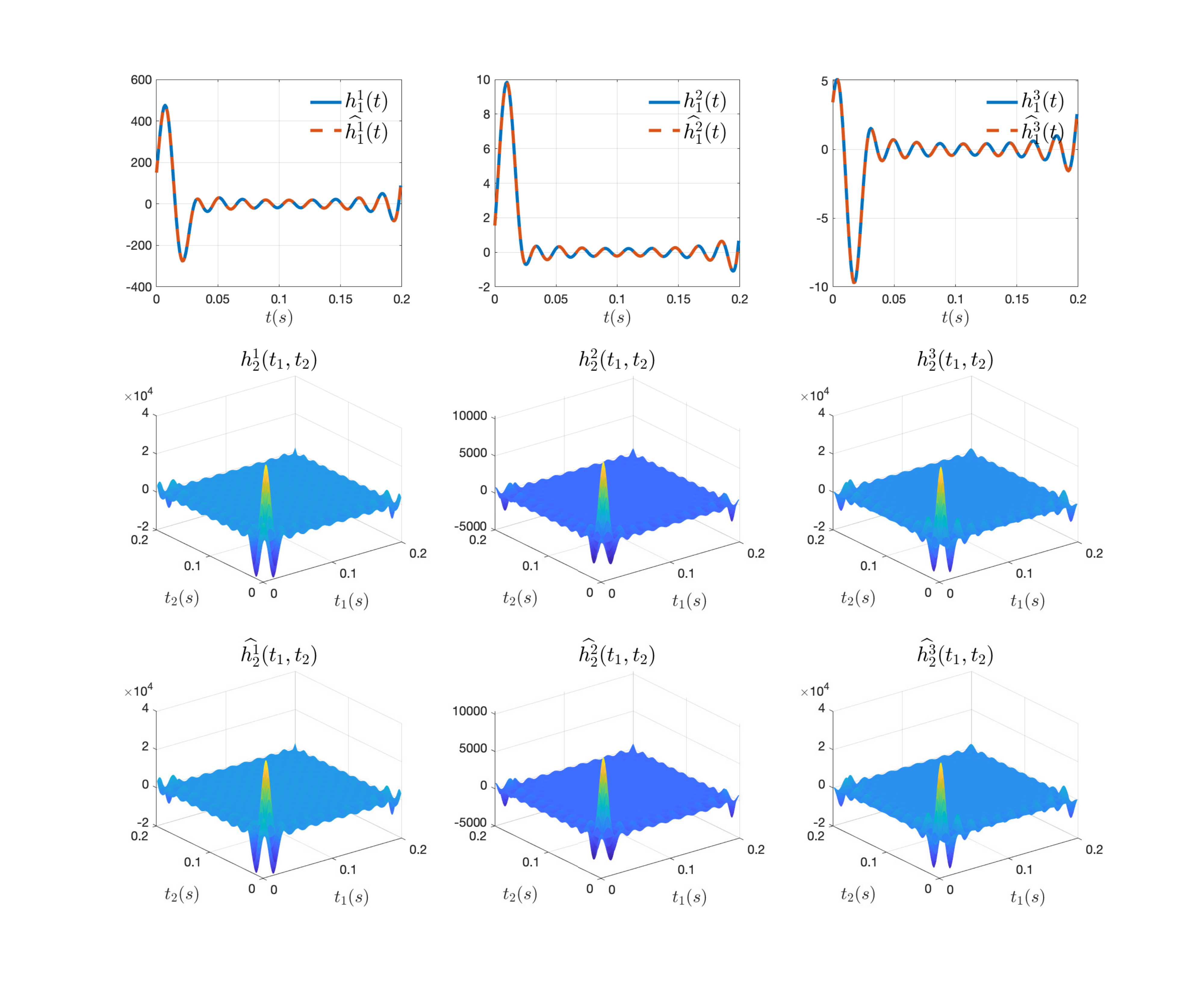}
\caption{Example of identification of a divisive normalization model. The SNRs of $\widehat{h^1_1}$, $\widehat{h^2_1}$, $\widehat{h^3_1}$, $\widehat{h_2^1}$, $\widehat{h^2_2}$ and $\widehat{h_2^3}$ were $57.4\;\mathrm{[dB]},$$\;\;56.19\;\mathrm{[dB]},$$\;\;47.01\;\mathrm{[dB]},$$ \;\;57.51\;\mathrm{[dB]}$, $57.52\;\mathrm{[dB]}$ and $57.51\;\mathrm{[dB]}$, respectively.
}
\label{fig:iden_norm}
\end{figure}
\end{example}

\begin{example}
Here we 
identify a biophysical model of the drosophila photoreceptor. 
A detailed biophysical model of Drosophila photoreceptors was given in\cite{SPB12, lazar_parallel_2015}. The photoreceptor model consists of 30,000 microvilli. The photon absorption in each microvillus is described by a Poisson process whose rate is proportional to the number of photons per microvillus incident on the ommatidium. Photon absorption leads to a transduction process governed by a cascade of chemical reactions. The entire transduction process described by 13 successive differential equations and is given in \cite{lazar_parallel_2015}.
The total number of equations of the photoreceptor model is $390,000$.

We presented the model mentioned above with a long stimulus consisting of random bandlimited fluctuations on top of different mean levels of light intensities. For each level, we used the first $3\%$ of the stimulus for identification and applied our identification framework on the resultant transduction current. 
The bandwidth of the input and output spaces was limited to $20$ Hz.
The identified filters are depicted in Figure~\ref{fig:pr_filters}. We additionally trained a model without normalization (\textit{i.e.}, with only $\mathcal{T}^1$); the outputs of the original biophysical model and the two identified models are shown in Figure~\ref{fig:pr_pred_out}. Without normalization, the identified model does not predict the output well across different light intensities. The SNR of the predicted output on training data for the normalization model was about $33 ~[dB]$, while the SNR of the predicted output on training data for the model without normalization was about
$19 ~[dB]$.

\begin{figure}[htbp]
	\centering
	\includegraphics[width=1.0\linewidth]{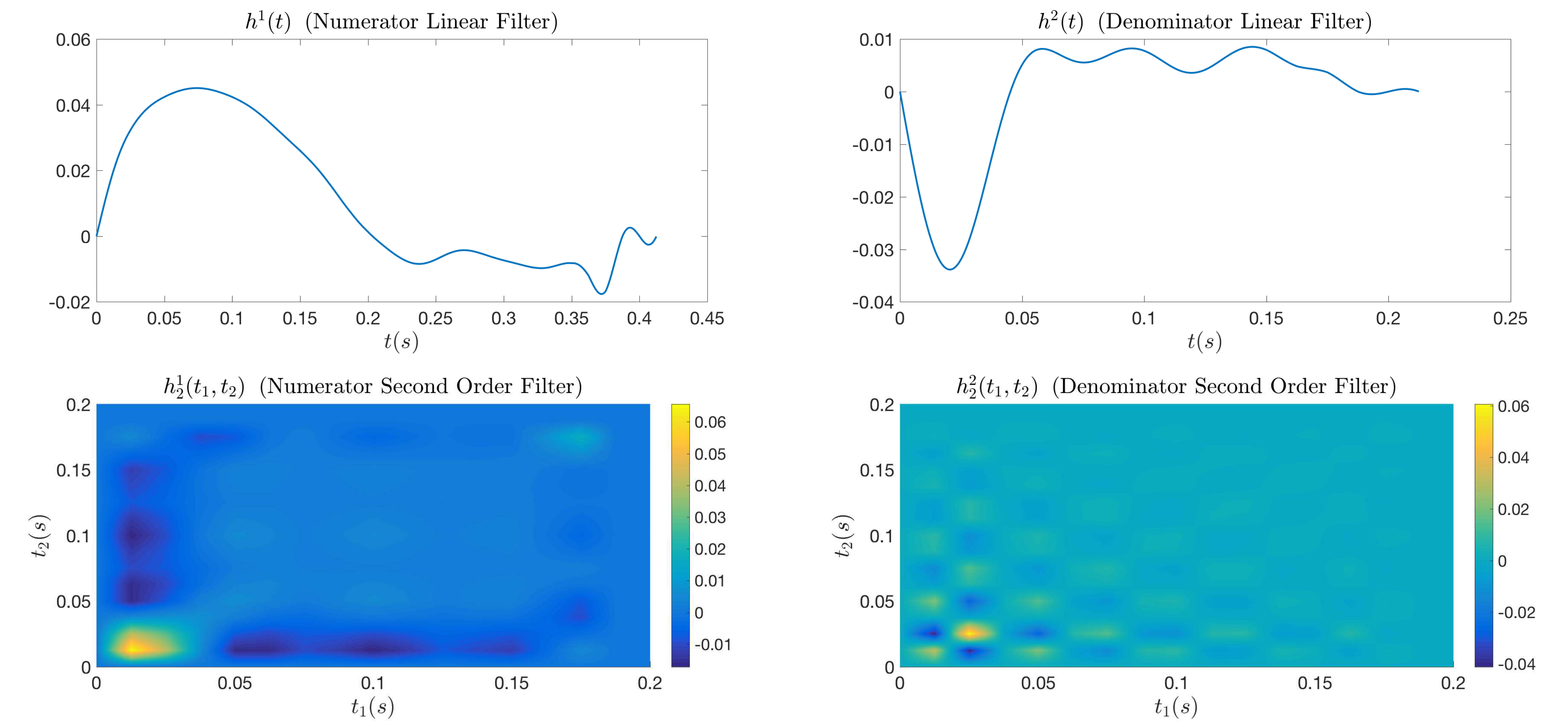}
	\caption{Identified filters of the biophysical photoreceptor model given in Example~\ref{ex:photoreceptor}.}
	\label{fig:pr_filters}
\end{figure}

\begin{figure}[t!]
	\centering
	\includegraphics[width=0.90\linewidth]{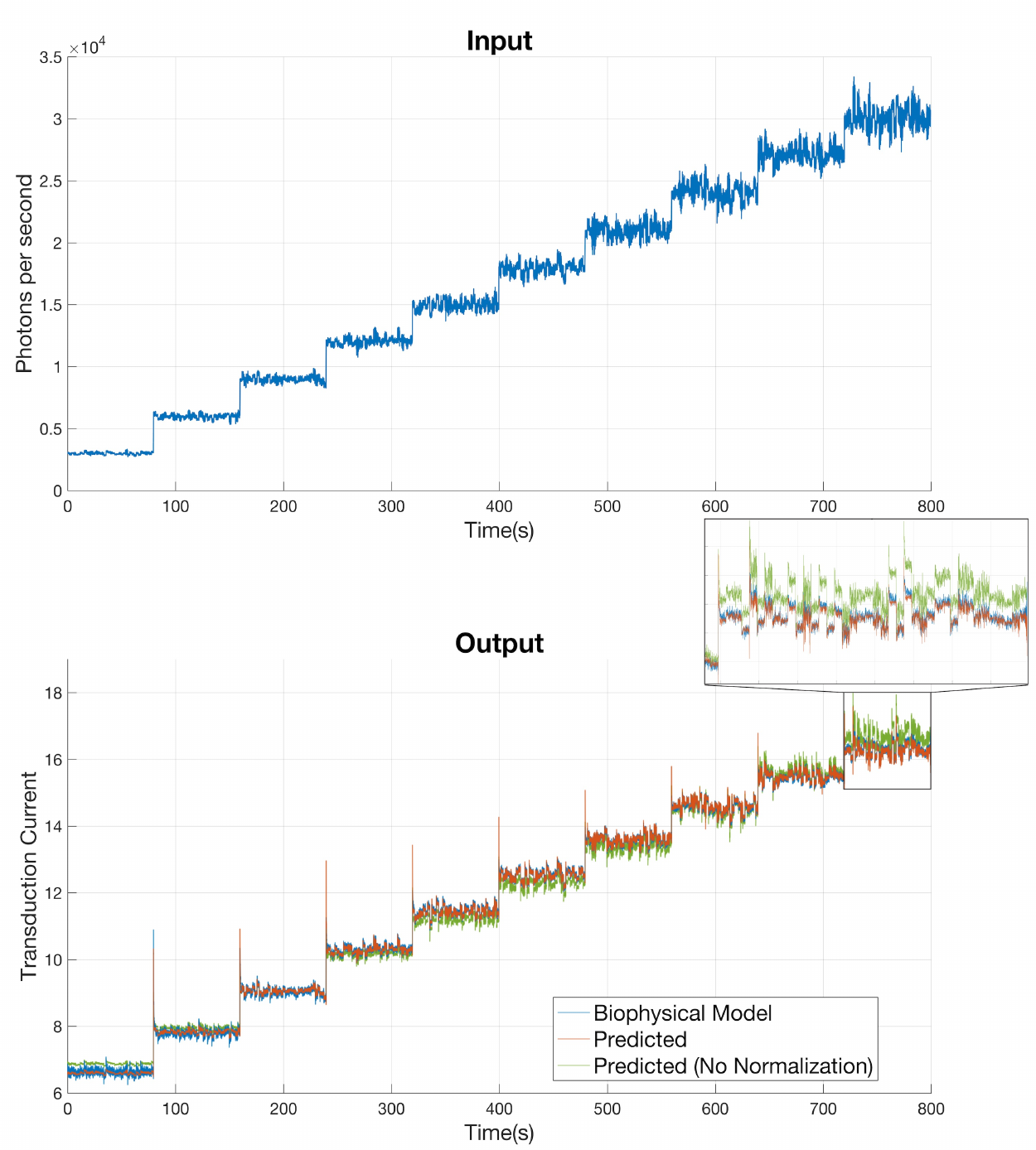}
	\caption{Comparison of the output of the biophysical photoreceptor model (blue) with the predicted output of the normalization model (red) and predicted output of a model without normalization, i.e., first and second order numerator filters only (green). Top panel depicts the input (photons per second).}
	\label{fig:pr_pred_out}
\end{figure}
\label{ex:photoreceptor}
\end{example}

\section{Sparse Identification of Spatio-Temporal DNPs}
\label{sec:space-time}

In  what follows we derive sparse identification algorithms for the components of spatio-temporal DNPs.

Given the spatio-temporal divisive normalization processor
depicted in Fig. \ref{fig:div_norm_st}, we are interested in identifying all the filters from input and output observations.
 We formulate an optimization problem, which achieves such identification, with high fidelity and with a relatively small number of measurements.

\subsection{Deriving the Sparse Identification Algorithm for Spatio-Temporal DNPs}
\label{sec:lateral_iden}

Here, we make the assumption that the filters $h^{i4}(t), i = 1,2,\cdots,N$, and $h_2^{ij4}(t_1, t_2)$, $i,j = 1,2,\cdots,N$, are followed by the LPF with bandwidth $\Omega^o$ and that the output is sampled at a rate of $2f_{max}$, where $f_{max} = \frac{\Omega}{2\pi}$ is the maximum of the bandwidth of the filters $h^{i4}(t)$ and $h_2^{ij4}(t_1, t_2)$. By abuse of notation, we will use $v^i(t)$ to denote the low-passed version of the actual output. Note that, based upon the assumptions on the bandlimited nature of the feedback filters acting on the output, the responses of these filters to the low-passed outputs will be the same as the responses to the actual outputs. 

We present $M$ trials where input stimuli $u^{nm}$ are chosen to be elements of the space of trigonometric polynomials $\hilbert_1$ for $n=1,\;\; \cdots,\;\;N,\;\;\;m=1,\;\;\cdots.\;\;M$.
We project the outputs $v^{nm}(t)$
on the Hilbert space of trigonometric polynomials $\hilbert_1^o$, i.e.,
$\mathcal{P}_1^o v^{nm} (s)$.
Note that $\mathcal{P}_1^o v^{nm} (s)$ is approximately $v^{nm} (s)$ for large values of $\Omega^o$ \cite{Martin2007}. Further we assume that $v^{nm}, h^1_1, h^2_1, h^3_1, h^{i4}_1\in \mathbb{L}^2(\mathbb{D}), i = 1,2,\cdots,N$, the space of square integrable functions over domain $\mathbb{D}$ and 
$h_2^1, h_2^2, h_2^3, h_2^{ij4} \in   \mathbb{L}^2(\mathbb{D}^2), i,j = 1,2,\cdots,N$, the space of square integrable functions over domain $\mathbb{D}^2$.

We consider here the identification of the entire DNP circuit at once, for two reasons. First, as all channels are connected in the spatial domain through the MVP,
the inputs to the MVP are the outputs of the entire DNP circuit.
Therefore, all outputs are required to identify the MVP. 
Second, since $h^i_1, h^i_2, i=1,2,3$, are shared across all channels,
less trials are needed to identify these filters. 
We present the following 
\begin{lemma}
With $M$ trials presented to the spatio-temporal DNP, let the inputs $u^{nm}(t)$ and the outputs $v^{nm}(t)$ be sampled at times $(t_k), k = 1, \cdots, T$.
Then, 
for $i,j=1,2,\cdots,N$, we have the following equations
\begin{align}
b^1 &+ 
\left\langle h_{1}^{1}, \phi_{11}^{nmk} \right\rangle_{\hilbert_1} \!\!\!+
	 \left\langle h_{1}^{2} , \phi_{12}^{nmk} \right\rangle_{\hilbert_1} \!\!\!+ 
	 \left\langle h_{1}^{3} , \phi_{13}^{nmk} \right\rangle_{\hilbert_1^o} 
	 + \sum\limits_{i=1}^N \left\langle h_1^{i4} ,
	 \phi_{14}^{inmk}
	 \right\rangle_{\hilbert_1^o} +
	 \label{eq:norm_inner_prod_form_st}
\\
	 &+
	 \left\langle h_{2}^{1}, \phi_{21}^{nmk} \right\rangle_{\hilbert_2} \!\!\!+ 
	 \left\langle h_{2}^{2}, \phi_{22}^{nmk}  \right\rangle_{\hilbert_2} \!\!\!+
	 \left\langle h_{2}^{3}, \phi_{23}^{nmk} \right\rangle_{\hilbert_2^o}
	 + \sum\limits_{i=1}^N \sum\limits_{j=1}^N \left\langle h_2^{ij4} , 
	 \phi_{24}^{ijnmk}
	 \right\rangle_{\hilbert_2^o} = q^{nmk} ,
	 \nonumber
\end{align}
where the sampling functions $\phi_{11}^{nmk} \in \hilbert_1, \;\; \phi_{12}^{nmk}\in\mathcal{H}_1,
 \;\; \phi_{13}^{nmk}\in\mathcal{H}_1^o, \;\; 
 \phi_{14}^{inmk}\in\mathcal{H}_1^o, \;\;\phi_{21}^{nmk}\in\mathcal{H}_2, \;\; \phi_{22}^{nmk}\in\mathcal{H}_2, \:\:
 \phi_{23}^{nmk}\in\mathcal{H}_2^o \mbox{ and } \phi_{24}^{ijnmk}\in\mathcal{H}_2^o$ are given by
\begin{align}
\phi_{11}^{nmk} (t) &= u^{nm}_1(t_k-t) , \nonumber\\
\phi_{12}^{nmk} (t)  &=  -q^{nm}(t_k) u^{nm}_1(t_k-t)  , \nonumber\\
\phi_{13}^{nmk} (t)  &=  -q^{nm}(t_k) (\mathcal{P}_1^o v^{nm}_1) (t_k-t)  , \nonumber\\
\phi_{14}^{inmk} (t)  &=  -q^{nm}(t_k) (\mathcal{P}_1^o v^{im}_1) (t_k-t)  ,
\label{eq:paxes2} \\
\phi_{21}^{nmk} (t, s)  &= u^{nm}_1(t_k-t)u^{nm}_1(t_k-s) ,  \nonumber\\
\phi_{22}^{nmk} (t, s)  &= -q^{nm}(t_k) u^{nm}_1(t_k-t)u^{nm}_1(t_k-s) , \nonumber\\
\phi_{23}^{nmk} (t, s)  &= -q^{nm}(t_k) (\mathcal{P}_1^o v^{nm}_1) (t_k-t)(\mathcal{P}_1^o v^{nm}_1)(t_k-s) ,\nonumber\\
\phi_{24}^{ijnmk} (t, s)  &= -q^{nm}(t_k) (\mathcal{P}_1^o v^{im}_1) (t_k-t) (\mathcal{P}_1^o v^{jm}_1) (t_k-s) ,
\nonumber
\end{align} 
and $q_k^{nm} =  v^{nm}_1(t_k)$ for all $m=1,2,...,M$, $k=1,2,...,T$
and $i,j,n=1,2,...,N$.
\label{thm:lateral_iden}
\end{lemma}
\proof See Appendix~\ref{sec:proof2}.

\begin{remark}
Theorem~\ref{thm:lateral_iden} suggests that identifying the lateral divisive normalization model is equivalent to solving a generalized sampling problem with noisy measurements. It also suggests that the output needs to be sampled at a high enough rate, and that the choice of the Hilbert space used to reconstruct the feedback filters is critical since incorrect choices for these parameters can negatively affect the identification by introducing \lq noise\rq ~in the measurements.
\end{remark}

We now present the following algorithm to identify the model that exploits the low-rank constraints imposed on the quadratic filters.

 \begin{lemma}
Equation  \eqref{eq:norm_inner_prod_form_st} can be expressed in matrix form as
\begin{equation}
\mathbf{c}_1^\trans \boldsymbol\Phi^{nmk} 
+ \operatorname{Tr}\left(\mathbf{C}_2^\herm \boldsymbol\Xi^{nmk}\right) = q^{nmk},\end{equation}
for all $n=1,2,\cdots,N$, $m=1,2,\cdots,M$ and $k=1,2,\cdots,T$,
where
\begin{align}
& \mbox{Measurements}  \leftarrow  q^{mk} \quad (scalar) \\
& \mbox{Unknowns}  \leftarrow  \left\{
\begin{array}{l}
\mathbf{c}_1 = \begin{bmatrix}
 				b^1 &
 				(\mathbf{h}^1_1)^\trans &
 				(\mathbf{h}^2_1)^\trans &
				(\mathbf{h}^3_1)^\trans &
				(\mathbf{h}^{14}_1)^\trans &
				(\mathbf{h}^{24}_1)^\trans &
				\cdots &
				(\mathbf{h}^{N4}_1)^\trans 
 \end{bmatrix}^\trans_{1 \times \left(4L+2(N+1)L^o+N+4)\right)}, \\
\mathbf{C}_2 = \begin{bmatrix}
(\mathbf{H}_2^1)^\trans & \!\!(\mathbf{H}_2^2)^\trans & \!\!\!\! \mathbf{0} & \!\!\!\!\mathbf{0} & \!\! \mathbf{0} & \!\! \!\! \cdots & \!\!\!\!\mathbf{0} \\
\mathbf{0} & \!\!\!\!\mathbf{0} & \!\!(\mathbf{H}_2^3)^\trans & (\mathbf{H}_2^{114})^\trans & (\mathbf{H}_2^{124})^\trans & \!\!\!\! \cdots & \!\!\!\!(\mathbf{H}_2^{NN4})^\trans
 \end{bmatrix}^\trans_{\left(2L+2L^o+2\right)\times \left(4L+2(N^2+1)L^o+N^2+3)\right)}
\end{array}
\right. \label{eq:c1_c2}\\
& \begin{array}{c}
\mbox{Sampling} \\ \mbox{matrices}
\end{array} \leftarrow  \left\{
\begin{array}{c}
\boldsymbol\Phi^{nmk} = \begin{bmatrix}
1 \\ \mathbf{u}^{nmk} \\ -q^{nmk}\mathbf{u}^{nmk} \\
-q^{nmk} \mathbf{v}^{nmk} \\
-q^{nmk} \mathbf{v}^{1mk} \\ 
-q^{nmk} \mathbf{v}^{2mk} \\ 
\vdots \\
-q^{nmk} \mathbf{v}^{Nmk}
\end{bmatrix}_{\left(4L+2(N+1)L^o+N+4)\right) \times 1} \\
\boldsymbol\Xi^{nmk} = 
\begin{bmatrix}
\mathbf{U}_2^{nmk} & \mathbf{0} \\
-q^{nmk}\mathbf{U}^{nmk} & \mathbf{0} \\
\mathbf{0} & -q^{nmk} \mathbf{V}_2^{nnmk} \\
\mathbf{0} & -q^{nmk} \mathbf{V}_2^{11mk} \\
\mathbf{0} & -q^{nmk} \mathbf{V}_2^{12mk} \\
\vdots & \vdots\\
\mathbf{0} & -q^{nmk} \mathbf{V}_2^{NNmk}
\end{bmatrix}_{
\left(4L+2(N^2+1)L^o+N^2+3)\right) \times\left(2L+2L^o+2\right)}
\end{array}
\right.
%    \end{array}
\end{align}.
\label{lem:lem2}
\end{lemma}

We provide the following algorithm to identify $b_1$, $\mathbf{h}^i_1, i= 1,2,3$,
$\mathbf{h}^{i4}_1, i= 1,2,\cdots,N$, $\mathbf{H}^i_2,i =1,2,3$, and $\mathbf{H}^{ij4}_2,i,j =1,2,\cdots,N$.

Again, we assume that all the second-order filters have sparse structures akin to \eqref{eq:h2_gen_form}.

\begin{algorithmp}
Let $\widehat{\mathbf{c}}_1$ and $\widehat{\mathbf{C}}_2$ be the solution to the following optimization problem 
\begin{align}
& \underset{\mathbf{c_1}, \;\mathbf{C_2},  \; \boldsymbol\varepsilon}{\mathrm{min}}
& & \| \mathbf{C_2} \|_* +\lambda_1 \|\mathbf{c_1}\|_2 + \lambda_2 \|\boldsymbol\varepsilon\|_2 \\
& \quad\mathrm{s.t}
&& \mathbf{c}_1^\trans \boldsymbol\Phi^{nmk} 
+ \operatorname{Tr}\left(\mathbf{C}_2^\herm \boldsymbol\Xi^{nmk}\right) = q^{nmk} 
+ \varepsilon^{(n-1)TM+(m-1)T+k}, \nonumber \\
&&&\quad\quad\quad\quad\quad\quad\quad\quad\quad\quad n=1,\cdots,N,m = 1,\cdots,M, k=1,\cdots,T, \label{eq:constraint1_st} \\
&&& \mathbf{1}^\trans{\boldsymbol\varepsilon} = {\bm 0} \label{eq:constraint2_st} \\
&&& \begin{bmatrix} 
\mathbf{I}_{2\dim(\mathcal{H}_1) } 
& \mathbf{0}
\end{bmatrix}
\mathbf{C_2}
\begin{bmatrix}
\mathbf{0} \\
\mathbf{I}_{\dim(\mathcal{H}_1^o)} 
\end{bmatrix}    =  \mathbf{0}, \label{eq:constraint3_st}  \\  %top-right 0
&&& \begin{bmatrix} 
 \mathbf{0}
& \mathbf{I}_{\dim(\mathcal{H}_1^0) } 
\end{bmatrix}
\mathbf{C_2}
\begin{bmatrix}
\mathbf{I}_{\dim(\mathcal{H}_1)} \\
\mathbf{0}
\end{bmatrix}    =  \mathbf{0},  \label{eq:constraint4_st} \\% bottom-right zero
&&&\begin{bmatrix} 
\mathbf{I}_{\dim(\mathcal{H}_1)}
& \mathbf{0}
\end{bmatrix}
\mathbf{C_2}
\begin{bmatrix}
\mathbf{I}_{\dim(\mathcal{H}_1)} \\
\mathbf{0} 
\end{bmatrix}    =
\begin{bmatrix}
\mathbf{I}_{\dim(\mathcal{H}_1)} \\
\mathbf{0}
\end{bmatrix}^\trans 
\mathbf{C_2}^\herm
\begin{bmatrix} 
\mathbf{I}_{\dim(\mathcal{H}_1)}
& \mathbf{0}
\end{bmatrix}^\trans\!\!,  \label{eq:constraint5_st} 
\\      %H^1 hermitian
&&& \begin{bmatrix} 
\mathbf{0}_{\dim(\mathcal{H}_1)} \!\!
& \mathbf{I}_{\dim(\mathcal{H}_1)} \!\! 
& \mathbf{0}
\end{bmatrix}
\mathbf{C_2}
\begin{bmatrix}
\mathbf{I}_{\dim(\mathcal{H}_1)} \\
\mathbf{0}
\end{bmatrix}    = 
\begin{bmatrix}
\mathbf{I}_{\dim(\mathcal{H}_1)} \\
\mathbf{0}
\end{bmatrix}^\trans 
\mathbf{C_2}^\herm
\begin{bmatrix} 
\mathbf{0}_{\dim(\mathcal{H}_1)} \!\!
& \mathbf{I}_{\dim(\mathcal{H}_1)} \!\!
& \mathbf{0}
\end{bmatrix}^\trans\!\!,  \label{eq:constraint6_st} \\  %H^2 hermitian
&&& 
\begin{bmatrix} 
\mathbf{0}
& \mathbf{I}_{\dim(\mathcal{H}_1^o)}  
\end{bmatrix}
\mathbf{C_2}
\begin{bmatrix}
\mathbf{0} \\
\mathbf{I}_{\dim(\mathcal{H}_1^o)} 
\end{bmatrix}    =  
 \begin{bmatrix}
\mathbf{0} \\
\mathbf{I}_{\dim(\mathcal{H}_1^o)} 
\end{bmatrix}^\trans 
\mathbf{C_2}^\herm
\begin{bmatrix} 
\mathbf{0}
& \mathbf{I}_{\dim(\mathcal{H}_1^o)}  
\end{bmatrix}^\trans\!\!. \label{eq:constraint7_st}
\end{align}
where $\lambda_1, \lambda_2$ are appropriately chosen hyperparameters
and 
$\varepsilon_{(n-1)TM+(m-1)T+k}  \in \real$ represent slack variables.
\label{al:lateral_iden}
\end{algorithmp}
The constraints in Algorithm~\ref{al:lateral_iden} are
similar to those in Algorithm~\ref{al:norm:iden}.
Note that $H^{ij4}$ are not constrained to be Hermitian.
This follows from the assumption that the MVP block may perform asymmetric processing on any pair of inputs to the block.
\begin{theorem}
The identified spatio-temporal divisive normalization is specified as $\hat{b}^1 = \hat{b}^1$ and
 \begin{align}
 \widehat{h_1^1} (t) &= \sum\limits_{l = -L}^{L} [\widehat{\mathbf{h}^1_1}]_{l+L+1} \cdot e_{l}(t), ~~~~~~~
 \widehat{h_1^2} (t) = \sum\limits_{l = -L}^{L} [\widehat{\mathbf{h}^2_1}]_{l+L+1} \cdot e_{l}(t), \\
 \widehat{h_1^3} (t) &= \sum\limits_{l = -L^o}^{L^o} [\widehat{\mathbf{h}^3_1}]_{l+L+1} \cdot e^o_{l}(t), ~~~~~
  \widehat{h_1^{i4}} (t) = \sum\limits_{l_= -L^o}^{L^o} [\widehat{\mathbf{h}^{i4}_1}]_{l+L+1} \cdot e^o_{l}(t),  \\
 \widehat{h_2^1} (t_1, t_2) &= \sum\limits_{l_{1} = -L}^{L}\sum\limits_{l_{2} = -L}^{L} [\widehat{\mathbf{H}^1_2}]_{l_{1}+L+1,L+1-l_2} \cdot e_{l_{1}}(t_1) e_{l_{2}}(t_2),\\
 \widehat{h_2^2} (t_1, t_2) &= \sum\limits_{l_{1} = -L}^{L}\sum\limits_{l_{2} = -L}^{L} [\widehat{\mathbf{H}^2_2}]_{l_{1}+L+1,L+1-l_{2}} \cdot e_{l_{1}}(t_1) e_{l_{2}}(t_2),\\
 \widehat{h_2^3} (t_1, t_2) & = \sum\limits_{l_{1} = -L^o}^{L^o}\sum\limits_{l_{2} = -L^o}^{L^o} [\widehat{\mathbf{H}^3_2}]_{l_{1}+L^o+1,L^o+1-l_{2}} \cdot e^o_{l_{1}}(t_1) e^o_{l_{2}}(t_2),\\
  \widehat{h_2^{ij4}} (t_1, t_2) & = \sum\limits_{l_{1} = -L^o}^{L^o}\sum\limits_{l_{2} = -L^o}^{L^o} [\widehat{\mathbf{H}^{ij4}_2}]_{l_{1}+L^o+1,L^o+1-l_{2}} \cdot e^o_{l_{1}}(t_1) e^o_{l_{2}}(t_2), i,j=1,2,\cdots,N ,
 \end{align}
where
\begin{align}
\begin{bmatrix}
 				\hat{b}^1 &
 				\widehat{\mathbf{h}^1_1}^\trans &
 				\widehat{\mathbf{h}^2_1}^\trans &
				\widehat{\mathbf{h}^3_1}^\trans &
				\widehat{\mathbf{h}^{14}_1}^\trans &
				\widehat{\mathbf{h}^{24}_1}^\trans &
				\cdots &
				\widehat{\mathbf{h}^{N4}_1 }^\trans
 \end{bmatrix}^\trans = \widehat{\mathbf{c}}_1 ,
 \end{align}
 and
\begin{align}
\begin{bmatrix}
\widehat{\mathbf{H}_2^1}^\trans& \widehat{\mathbf{H}_2^2}^\trans 
& - & - & - & \cdots & - \\
- & - & \widehat{\mathbf{H}_2^3}^\trans & \widehat{\mathbf{H}_2^{114}}^\trans  & \widehat{\mathbf{H}_2^{124}}^\trans  & \cdots & \widehat{\mathbf{H}_2^{NN4}}^\trans
 \end{bmatrix}^\trans = \widehat{\mathbf{C}}_2 .
 \label{eq:C2_st}
 \end{align}
\end{theorem}

\begin{remark}
	Assuming all the second-order kernels in the lateral divisive normalization model have rank $r$, the expected number of measurements for Algorithm~\ref{al:lateral_iden} to identify the model is of the order $\mathcal{O}\left(r \cdot \dim(\hilbert_1) + N^2r \cdot dim(\hilbert_1^o)\right)$. When $N$ is large, the $N^2$ factor may become prohibitive in identifying the model. Additional assumptions on $h^{ij4}_2$ may help mitigate this and maintain tractability of solving the identification problem. For example, with the assumption that $h^{ij4} = h^{14}_1$ if $i=j$ and $h^{ij4}_2 = h^{24}_1$ otherwise, the expected number of measurements required will  $\mathcal{O}\left(r \cdot \dim(\hilbert_1) + Nr \cdot dim(\hilbert_1^o)\right)$.
\end{remark}

\subsection{An Example of Sparse Identification of a Spatio-Temporal DNP}

We now present an example of identification obtained by Algorithm~\ref{al:lateral_iden}.
We demonstrate here that, in addition to the identification of the Volterra
kernels operating within each channel, the
MVP in the spatio-temporal DNP can be identified.
\begin{example}
\label{ex:st_dnp}
Here, we choose the DNP in Fig. \ref{fig:div_norm_st} with $N=4$, and 
\begin{equation}
h^1_1(t) = 25 t e^{-25t},
\end{equation}
%and the linear filters in VO-FF-ib blocks were chosen to be
\begin{equation}
h^2_1(t) = 25te^{-25t},
\end{equation}
%The linear filters in the MVO-FB block were chosen to be
\begin{equation}
h^{i4}_1(t) = e^{ -\frac{1}{4}(i-2)^2}(25-600t)e^{-25t},
\end{equation}
%Finally, the quadratic filters in the MVO-FB block were chosen to be
\begin{equation}
h^{ij4}_2(t_1,t_2) = 5,000 e^{ -\frac{1}{4}(i-2)^2}e^{ -\frac{1}{4}(j-2)^2} \cdot (25t_1e^{-25t_1})(25t_2e^{-25t_2}),
\end{equation}
and the other Volterra kernels are set to 0.
Note that Yiyin{, and $h^1_1$, $h^2_1$ are shared
across all channels. In addition,} $h^{ij4} = h^{ji4}$ in this case. 
We assumed knowledge of this symmetric structure of the model and adapted the identification algorithm accordingly and only identified 6 linear filters and 10 quadratic filters.

\begin{figure}
	\begin{minipage}{\textwidth}
		\begin{minipage}{0.65\textwidth}
			\vspace{0.1in}
				\centering
				\includegraphics[width=1.0\textwidth]{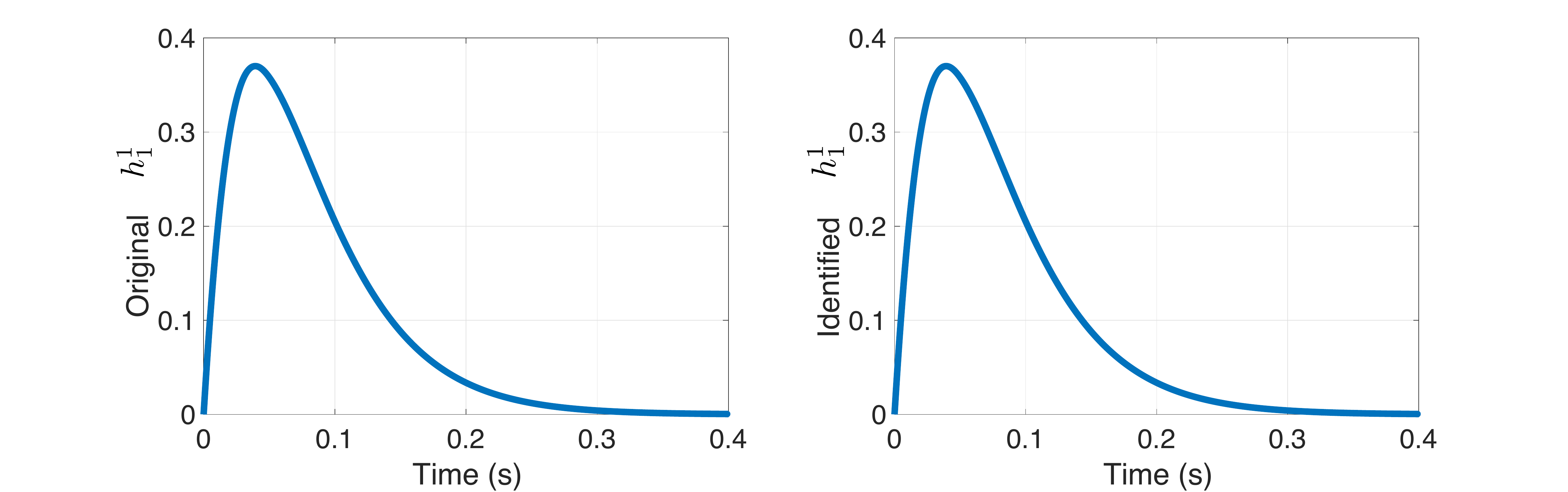}
				\centering
				\includegraphics[width=1.0\textwidth]{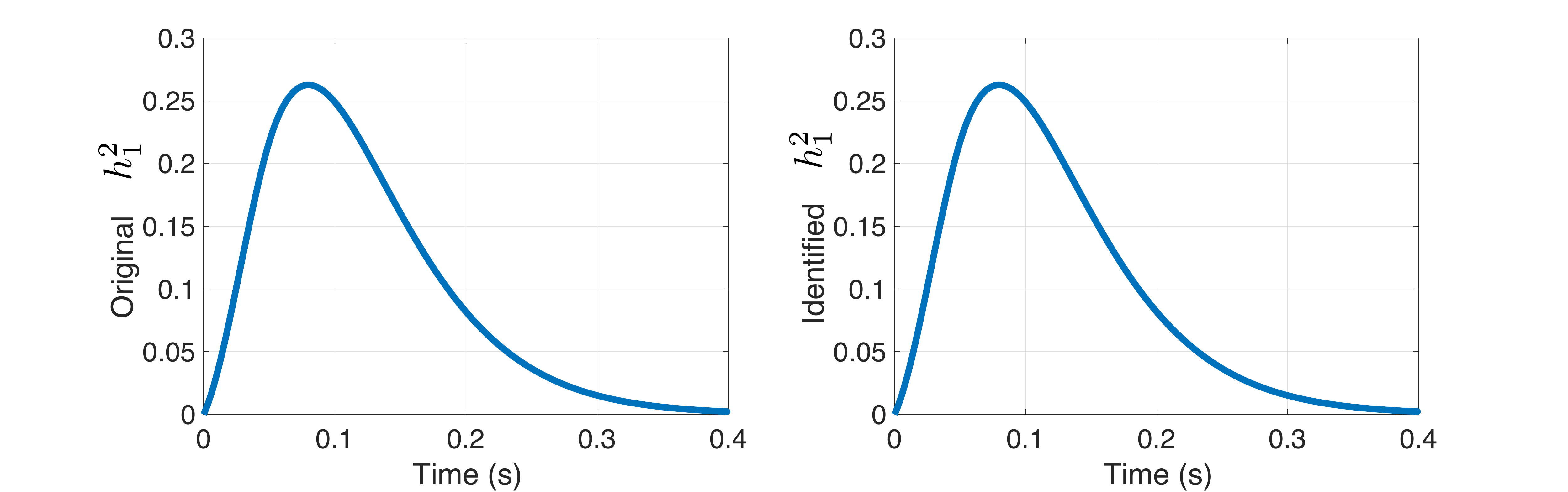}
				\centering
				\includegraphics[width=1.0\textwidth]{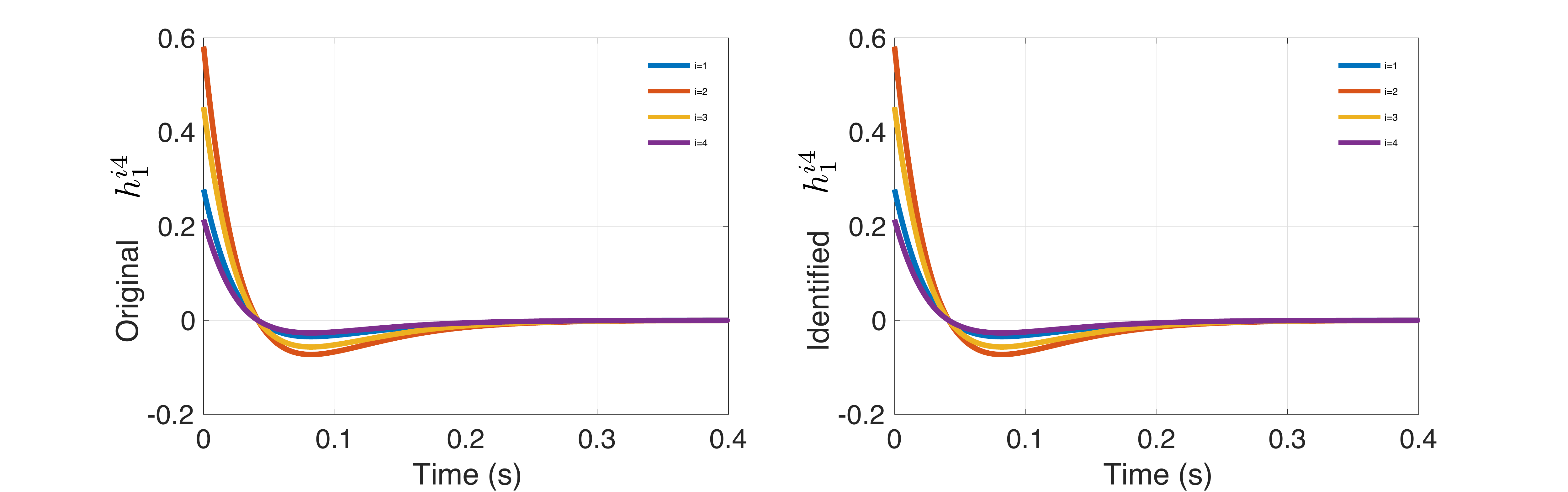}
%			\vspace{0.1in}
			\begin{center}{\footnotesize  First-order Filters}\end{center}
		\end{minipage}
		\begin{minipage}{0.33\textwidth}
			\begin{minipage}{0.4\textwidth}
				\begin{center}{\footnotesize Original}
				\end{center}
			\end{minipage}
			\begin{minipage}{0.4\textwidth}
				\begin{center}{\footnotesize Identified}
				\end{center}
			\end{minipage}
			\hspace{-0.2in}
			\centering
			\includegraphics[width=0.9\textwidth]{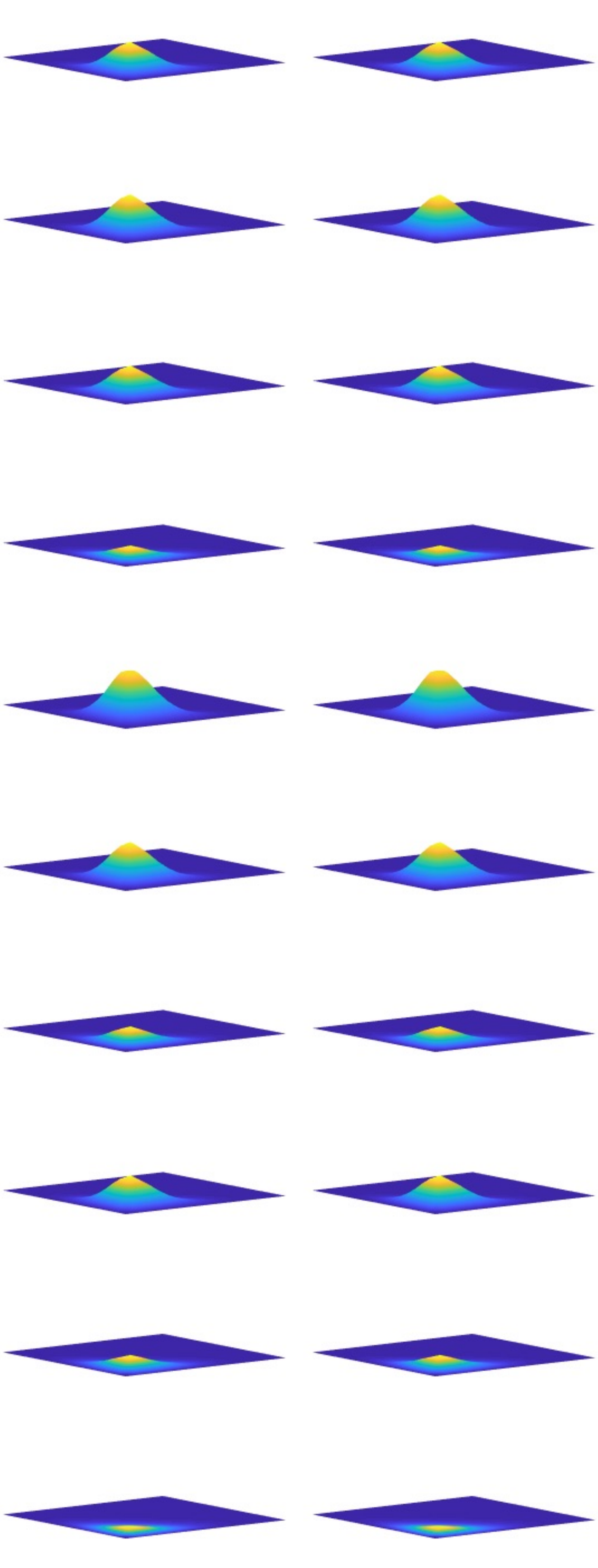}
			{\footnotesize MVP Second-order Kernels}
		\end{minipage}
 	\end{minipage}
 	\caption{Example of identification of the spatio-temporal DNP given in Example~\ref{ex:st_dnp}. (left) Identification of the first-order filters (from top to botom) $h_1^1, h_1^2$ and $h_{1}^{i4}, i=1,2,3,4$. (right) Identification of the second-order filters (from top to botom) $h_2^{ij4}$ with $i \leq j$ for $i=1,2,3,4$ and $j=1,2,3,4$.
}
 	\label{fig:lateral_iden_ex}
\end{figure}

We performed the identification in the Hilbert space of bandlimited functions with $\Omega = \Omega^o = 40\pi$ and we chose the same space for the input stimuli and for both the feedforward and feedback filters. We solved for the filters truncated to a period of $0.4s$, and thus there were $17$ coefficients to be identified for the linear filters and $17\times 17$ coefficients to be identified for each of the quadratic filters. A total of $1,116$ measurements ($279$ measurements from each cell) were used to perform the identification and the results are depicted in Figure~\ref{fig:lateral_iden_ex}. The average SNR of reconstruction across all filters was more than $150 ~[dB]$. Note that solving the generalized sampling problem directly for the same problem would have required at least $3,570$ measurements.
 \end{example}

\section{Discussion}
\label{sec:disc}

As already mentioned in the introduction, the photoreceptor/amacrine cell  layer
of the early vision system of the fruit fly rapidly adapts to visual stimuli whose intensity and contrast vary orders of magnitude both in space and time.

In this paper we presented a spatio-temporal divisive normalization processor
that models the transduction and the contrast gain control in the photoreceptor and amacrine cell layer
of the fruit fly.
It incorporates processing blocks that explicitly model the feedforword and the temporal feedback path of each photoreceptor and the
spatio-temporal  feedback from amacrine cells to photoreceptors.
We demonstrated that with some simple choice of parameters, the DNP response 
maintains the contrast of the input visual field across a large range of average spatial luminance values.

We  characterized the I/O of the spatio-temporal DNP
and highlighted the highly nonlinear behavior of the DNP  
in contrast gain control.
Despite the divisive nonlinearity, we 
provided an algorithm for the sparse identification of the entire DNP.
We showed that the identification of the
DNP can be interpreted as a generalized
sampling problem. 
More importantly, the sparse identification algorithm
does not suffer from the curse of dimensionality that would otherwise
require a large number of measurements that is quadratically related to the
dimension of the input and output spaces.

The DNP model opens a new avenue
for exploring and quantifying the highly nonlinear
nature of sensory processing. 
The DNP in Figure~\ref{fig:div_norm_st} can be further extended to allow the 
different transformations $\mathcal{T}^i, i=1,2,3$,
to incorporate spatio-temporal Volterra kernels, thereby making it more  versatile for modeling other types of sensory processing, including
(i) interactions between cones and horizontal cells in  vertebrate retinas \cite{VanLeeuwen2009},
(ii) channels/glomeruli in olfactory circuits and interactions between them through local neurons \cite{RFC10, Wilson2013}, and
(iii) cross-suppression and gain control in the auditory \cite{Rabinowitz2011, Natan2017}, and visual corticies \cite{allison_smith_bonds_2001, Geisler1992, Priebe2006}.

\newpage
\appendix

\section{Spatio-Temporal DNPs and Contrast Gain Control}
\label{sec:I/O_lateral} 

Here we characterize the I/O of simple DNPs stimulated with  three different inputs.
In the first example, we evaluate the response of a $1 \times 4$ DNP 
under different background light intensity levels.
In the second example, contrast gain control exerted by the amacrine cells is demonstrated with the same DNP.
In the third example, a DNP consisting of $16 \times 16$ DNPs tiling a
$1,536 \times 1,024$ visual field 
 is stimulated with a natural image taken at low, medium and high luminance values.
The DNP output is evaluated with and without the MVP block.

\begin{example}
Here, we consider a simple $1 \times 4$ DNP consisting of 4 photoreceptors
and a single amacrine cell receiving inputs from and providing feedback to
all 4 photoreceptors. The choices of the component filters of the DNP are as follows:
\begin{equation}
b_2+b_3+b_4 = 1,
\end{equation}
for the transformations $\mathcal{T}^1$ and $\mathcal{T}^2$ the kernels are:
\begin{equation}
h_{1}^{1}(t) = h_{1}^{2}(t) = \frac{t}{0.1} \operatorname{exp}\left(-\frac{t}{0.1}\right),
\end{equation}
\begin{equation}
h_{2}^{1}(t_1,t_2) = h_{2}^{2}(t_1,t_2) = 0.0001 h_{1}^{1}(t_1)h_{1}^{1}(t_2),
\end{equation}
for the transformation $\mathcal{T}^3$ the kernels are,
\begin{equation}
h_{1}^{3}(t) = 0,
\end{equation}
\begin{equation}
h_{2}^{3}(t_1,t_2) = 0,
\end{equation}
and for the transformation $\mathcal{L}^4$ the kernels are:
\begin{equation}
h^{i4}_{1}(t) = -10,000 \frac{t}{0.8}\operatorname{exp}\left(-\frac{t}{0.2}\right), i = 1,2,3,4,
\end{equation}
\begin{equation}
h^{ij4}_{2}(t_1,t_2) = \left\{
\begin{array}{cc} 
12,500\frac{t_1 t_2}{0.04} \operatorname{exp}\left(-\frac{t_1+t_2}{0.2}\right), & i = j,\\
0, & i \neq j,
\end{array} \right. i, j = 1,2,3,4.
\end{equation}
Here, the bandwidth of $\hilbert^o_1$ was chosen to be $\Omega^o = 10\cdot 2\pi$ rad/s.
The I/O of the fly's individual photoreceptors in steady state is described by a saturating
non-linearity similar in appearance to a sigmoid when plotted on a logarithmic scale \cite{LAUGHLIN1994}. 
Photoreceptors can deal with the orders of magnitude of the input stimuli while maintaining their 
output in a suitable range. In addition, the photoreceptors exhibit adaptation so that the sigmoidal curves 
shift to adjust to the mean local luminance value \cite{CH2012}.

We examine here the steady state response of the above DNP
under 4 different background light intensity levels.
At the start of each trial, all 4 photoreceptors are first adapted to the same background light intensity
level. One of the photoreceptors is then subject to an additional flash of different
light intensity with a duration of 2 seconds, while the inputs to the other 3 are kept at the same background level.
We observe the value of the steady state response of the photoreceptor that receives
the additional flash. 
Figure~\ref{fig:io_flashes} depicts the relationship  between the observed steady
state response and the light intensity of the flash, at the 4 background intensity levels. It demonstrates that the response of the DNP is background-dependent,
and the overall slope and range are similar across different background levels.
With the MVP block, the spatio-temporal DNP can reproduce responses
of photoreceptors observed in experimental settings \cite{CH2012, ML81}.

\begin{figure}[t!]
\centering
\includegraphics[width=0.5\textwidth]{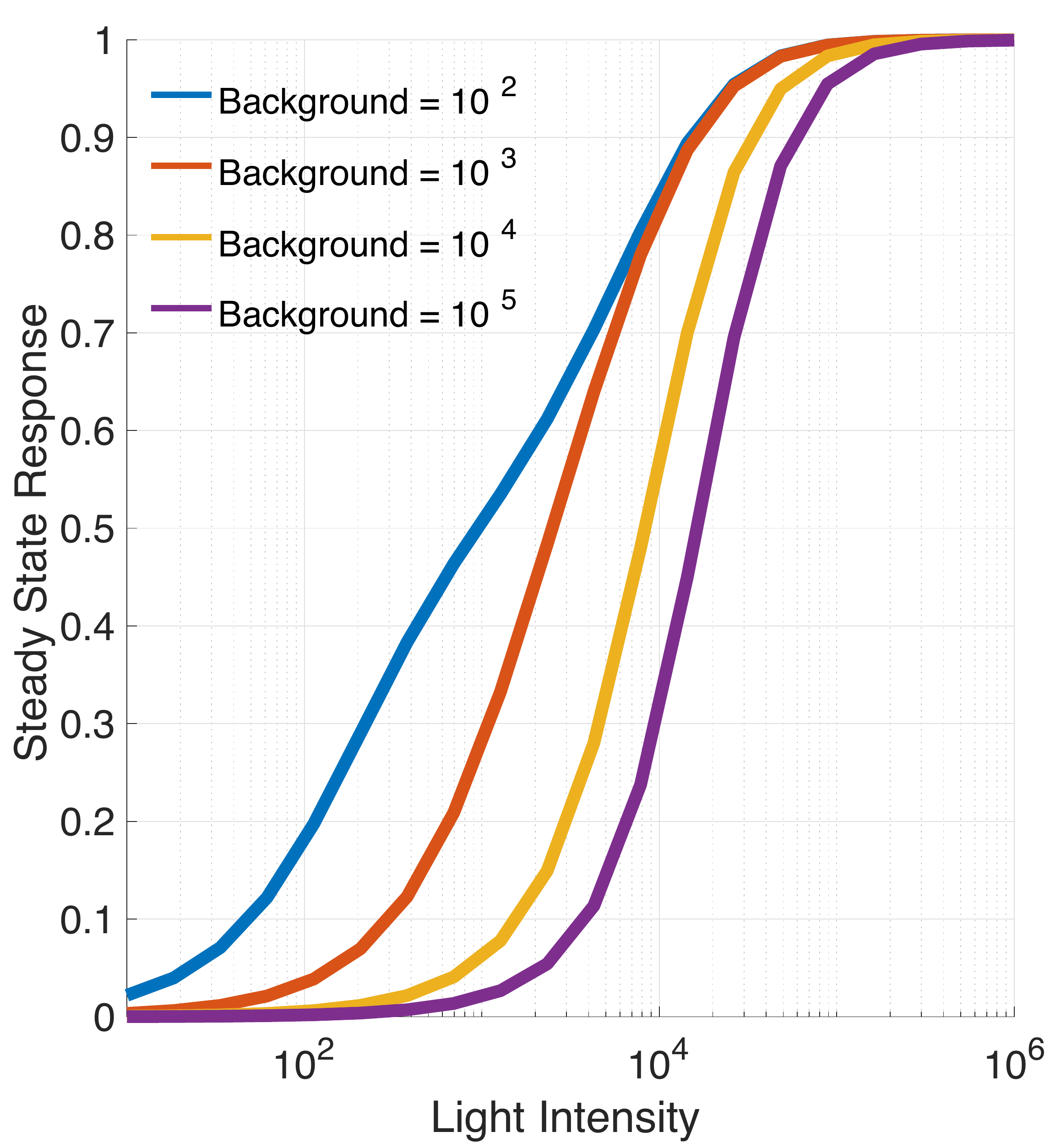}
\caption{The spatio-temporal DNP model can exhibit adaptation to local luminance. 
A DNP with 4 photoreceptors and an amacrine cell (see Example 1 for details) 
is adapted to a background intensity level at the beginning of each trial.
One of the photoreceptors is then provided a 2-second long flash of light,
while the inputs to the other photoreceptors are kept at the background level.
The relationship between the steady state response of the photoreceptor and
the light intensity of the flash is shown for each of the background levels.
}
\label{fig:io_flashes}
\end{figure}

\end{example}

Since all the photoreceptors exhibit a sigmoid like non-linearity, the output of the 
retina will be constrained in a suitable
current range that can be processed by 
postsynaptic neurons in the lamina. 
However, without adaptation to mean local luminance, the saturating nature of the
non-linearities leads to a loss of spatial contrast. The spatial contrast is preserved 
by spatial gain control or adaptation modeled here with the MVP block. 

\begin{example}
Here, the I/O of DNPs with and without the MVP block is evaluated.
Using the same DNP as in the example above, we
stimulated the DNP with 25 ``images"
with a resolution of $1\times4$ pixels.
Each image has a different average luminance and root mean square (RMS) contrast.
The RMS contrast is defined as the standard deviation of pixel intensities
normalized by the mean i.e.,
\begin{equation}
C_{rms} = \dfrac{\sqrt{\frac{1}{N} \sum_{i=1}^{N} \left(u^i - \overline{u}\right)^2}}{\overline{u}},
\end{equation}
where
\begin{equation}
\overline{u} = \frac{1}{N} \sum_{i=1}^{N} u^i
\end{equation}
and $N=4$.
These images are shown in the ``input" block in Figure~\ref{fig:lateral_bar_io},
with each bar represents the input intensity to one photoreceptor.
Note that the pixels are extended
along the y-axis for a quick visual examination. 

 \begin{figure}[tbp]
	\centering
	\includegraphics[width=\textwidth]{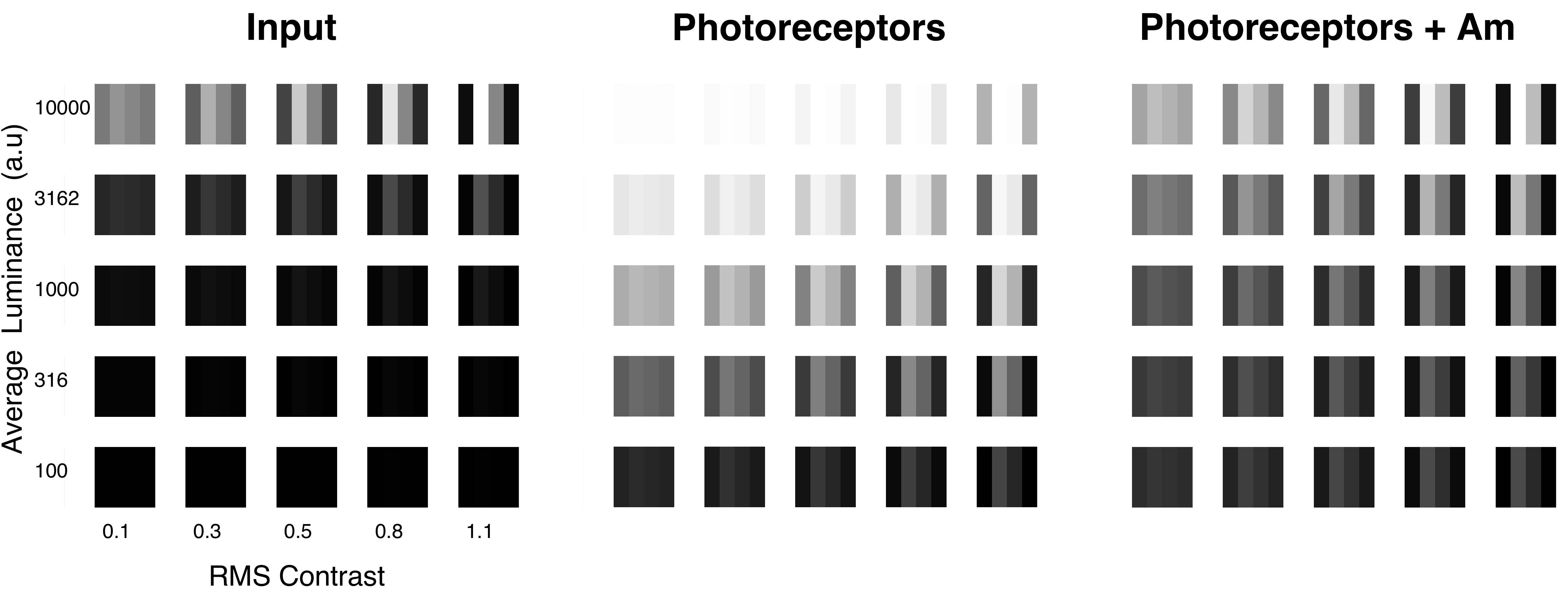}
	\caption{Steady State I/O visualization for the DNP considered in Example 1 . Left column - input stimuli comprising bar gratings at various luminances and RMS contrasts. Middle column - Responses of the DNP without the MVP block. Right Column - Responses of the DNP with the MVP block.}
	\label{fig:lateral_bar_io}
\end{figure}

In the ``Photoreceptors" block in Figure~\ref{fig:lateral_bar_io},
the steady state responses of the DNP without the MVP block to
the respective inputs are shown. Here pure black represents
a response of 0 and white 1. This can be interpreted as a circuit
in which the reciprocal connections between photoreceptors and amacrine cells 
are blocked.

In the ``Photoreceptors $+$ Am" block in Figure~\ref{fig:lateral_bar_io},
the steady state responses of the full DNP to their respective inputs
are shown. Comparing the ``Photoreceptors"  and ``Photoreceptors $+$ Am" block,
the responses of the DNP without feedback are washed out, or exhibit low contrast, particularly at three of the four corners of the $5\times 5$ image array,
\textit{i.e.}, when either the luminance or contrast is too high or too low.
By contrast, the individual bars in the response of the full DNP model
is more readily discernible across several orders of magnitude of luminance.

\begin{figure}[htbp]
	\centering
	\includegraphics[width=1.0\textwidth]{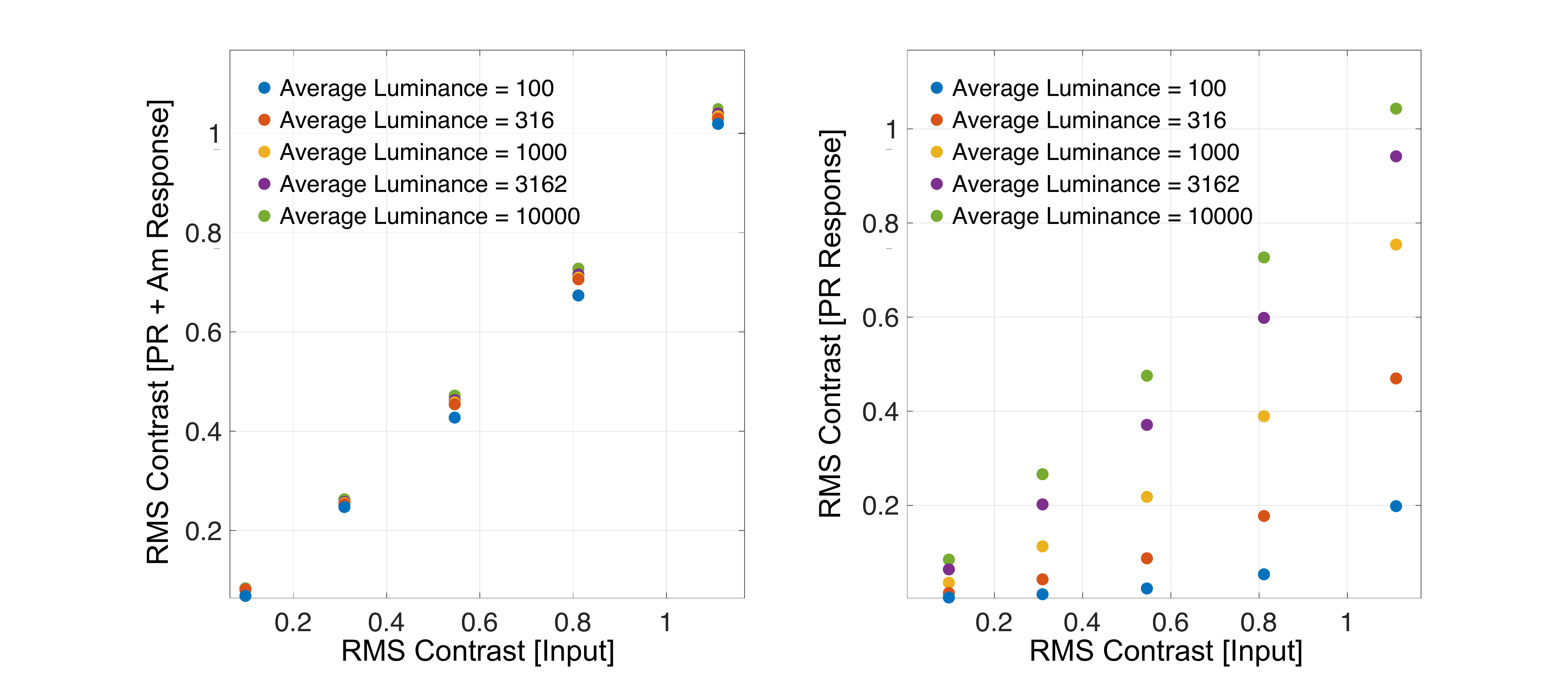}
	\caption{RMS contrast of the input images in Figure~\ref{fig:lateral_bar_io} are plotted against the RMS contrasts of the responses with (left) and without (right) the MVP block.}
	\label{fig:contrast_contrast}
\end{figure}

We note that a saturating non-linearity 
can maintain its output in a constrained range even when the input varies by orders of magnitude.
However, the non-linearity may lead to a loss of spatial contrast.
In contrast the DNP, constrains its output to a suitable range while maintaining  the spatial contrast.
This is demonstrated in Figure~\ref{fig:contrast_contrast}, where the 
RMS contrast of the input images in Figure~\ref{fig:lateral_bar_io} are plotted 
against the RMS contrasts of the responses for both cases - with and without 
MVP in the DNP.  Spatial contrast, arguably, is an important feature of 
the image that should be preserved or even enhanced for subsequent stages 
of extraction of ethologically relevant information from the visual stimulus.
\end{example}

\begin{example}
Here we apply a full-scale DNP model to a natural image.
The image is taken in raw
format so that its pixel values are proportional to the light intensity each pixel
is exposed to \cite{HS98}.
The resolution of the image is $1,536\times 1,024$.
The image is first divided into $16\times 16$ blocks with a 4 pixel overlap in
each direction.
A DNP is assigned to each block, and
the filters in the DNP are designed as follows:
\begin{equation}
b_2+b_3+b_4 = 1,
\end{equation}
for the transformations $\mathcal{T}^1$ and $\mathcal{T}^2$ the kernels are:
\begin{equation}
h_{1}^{1}(t) = h_{1}^{2}(t) = \frac{t}{0.04} \operatorname{exp}\left(-\frac{t}{0.04}\right),
\end{equation}
\begin{equation}
h_{2}^{1}(t_1,t_2) = h_{2}^{2}(t_1,t_2) = 0.0001 h_{1}^{1}(t_1)h_{1}^{1}(t_2),
\end{equation}
for the transformation $\mathcal{T}^3$ the kernels are:
\begin{equation}
h_{1}^{3}(t) = 0,
\end{equation}
\begin{equation}
h_{2}^{3}(t_1,t_2) = 0,
\end{equation}
and for the transformation
$\mathcal{L}^4$ the kernels are:
\begin{equation}
h^{i4}_{1}(t) = -20 \frac{t}{0.04}\operatorname{exp}\left(-\frac{t}{0.04}\right) \cdot \frac{1}{\sqrt{2\pi\cdot16}}\operatorname{exp}\left( - \frac{(x^i-x^0)^2+(y^i-y^0)^2}{32}\right) , i = 1,2,3,4,
\end{equation}
\begin{equation}
h^{ij4}_{2}(t_1,t_2) = \left\{
\begin{array}{cc} 
100\frac{t_1 t_2}{0.04^2} \operatorname{exp}\left(-\frac{t_1+t_2}{0.04}\right) \cdot \sqrt{\frac{1}{\sqrt{2\pi\cdot16}}\operatorname{exp}\left( - \frac{(x^i-x^0)^2+(y^i-y^0)^2}{32}\right)} , & i = j,\\
0, & i \neq j,
\end{array} \right. i, j = 1,2,3,4,
\end{equation}
where $(x^i, y^i)$ is the coordinate of pixel $i$, and $(x^0, y^0)$ is the coordinate of
the center pixel in the $16\times 16$ block.
Here, the bandwidth of $\hilbert^o_1$ was chosen to be $\Omega^o = 40\cdot 2\pi$ rad/s.

\begin{sidewaysfigure}[htbp]

\begin{minipage}{\textwidth}
	\begin{minipage}{0.12\textwidth}
		\begin{center}
			\centering
			Input
		\end{center}
	\end{minipage}		
	\begin{minipage}{0.26\textwidth}
			\includegraphics[width=0.98\textwidth]{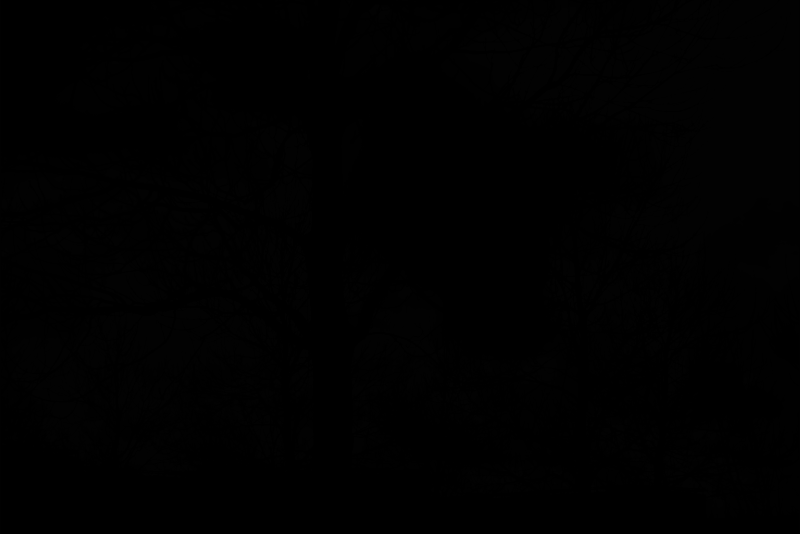}
	\end{minipage}		
	\begin{minipage}{0.26\textwidth}
			\includegraphics[width=0.98\textwidth]{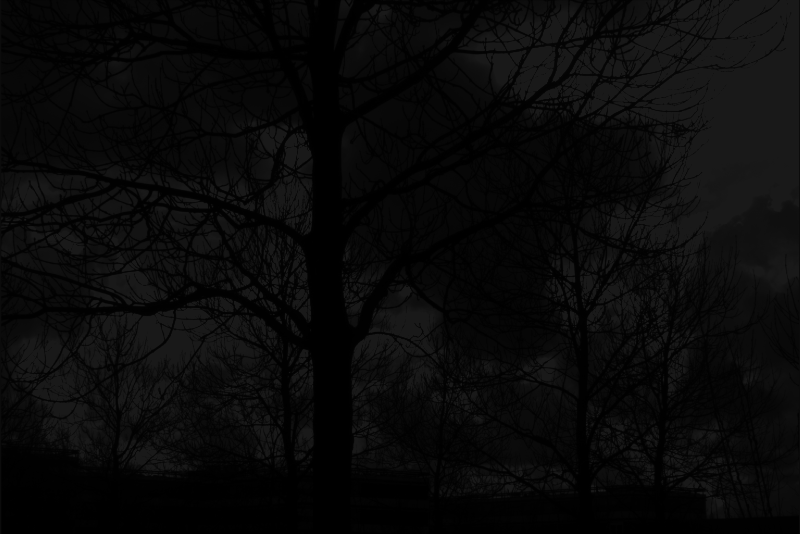}
	\end{minipage}	
	\begin{minipage}{0.26\textwidth}
			\includegraphics[width=0.98\textwidth]{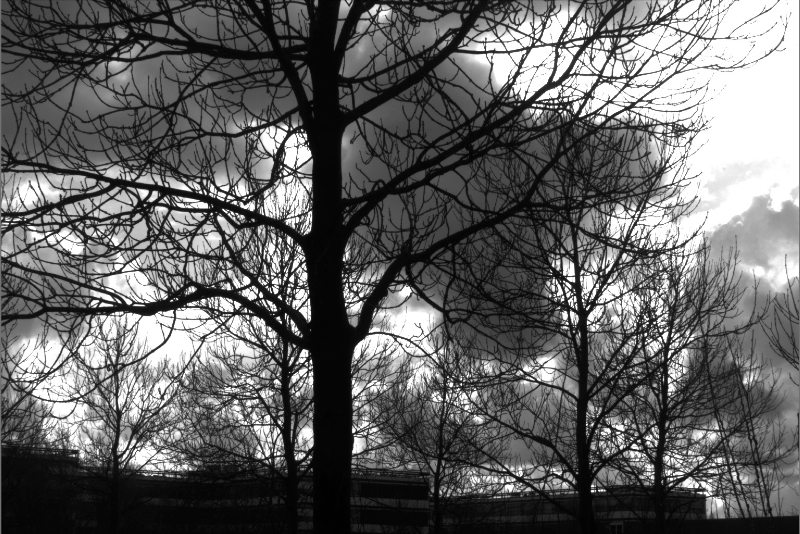}
	\end{minipage}	
	\begin{minipage}{0.01\textwidth}
		
	\end{minipage}
\end{minipage}
\begin{minipage}{\textwidth}
	\begin{minipage}{0.12\textwidth}
		\begin{center}
			 Log(Input)
		\end{center}
	\end{minipage}	
	\begin{minipage}{0.26\textwidth}
			\includegraphics[width=0.98\textwidth]{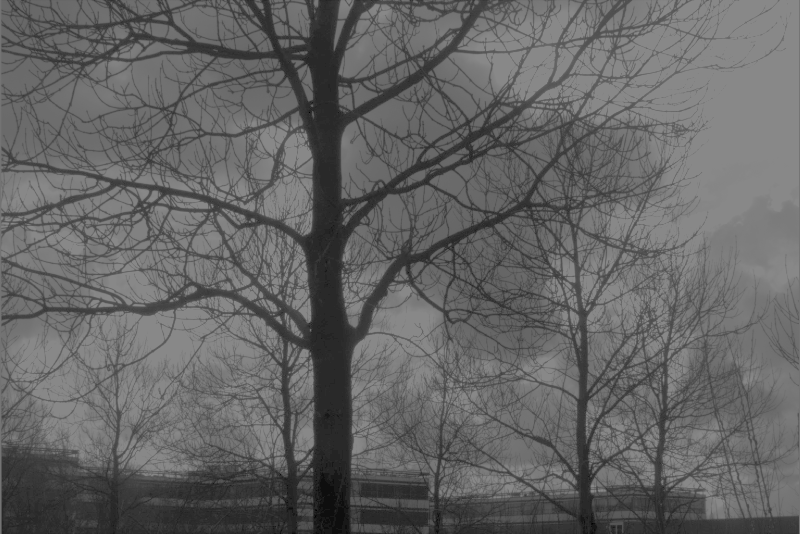}
	\end{minipage}	
	\begin{minipage}{0.26\textwidth}
			\includegraphics[width=0.98\textwidth]{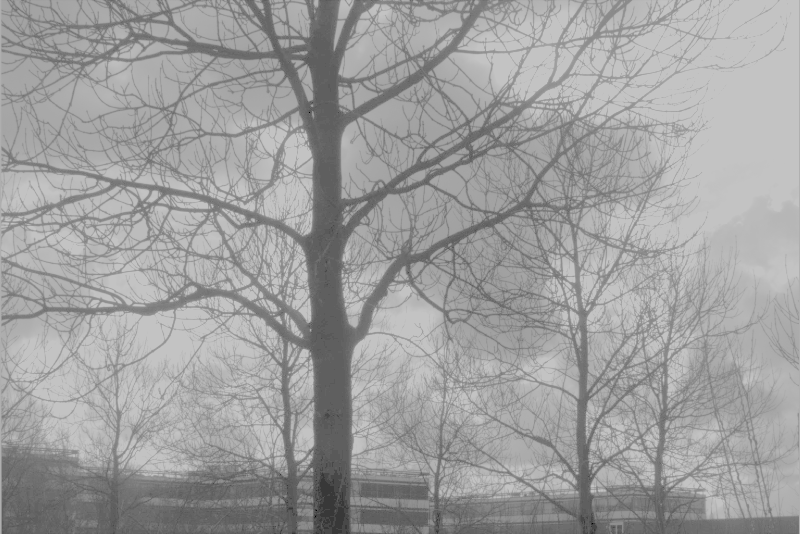}
	\end{minipage}	
	\begin{minipage}{0.26\textwidth}
			\includegraphics[width=0.98\textwidth]{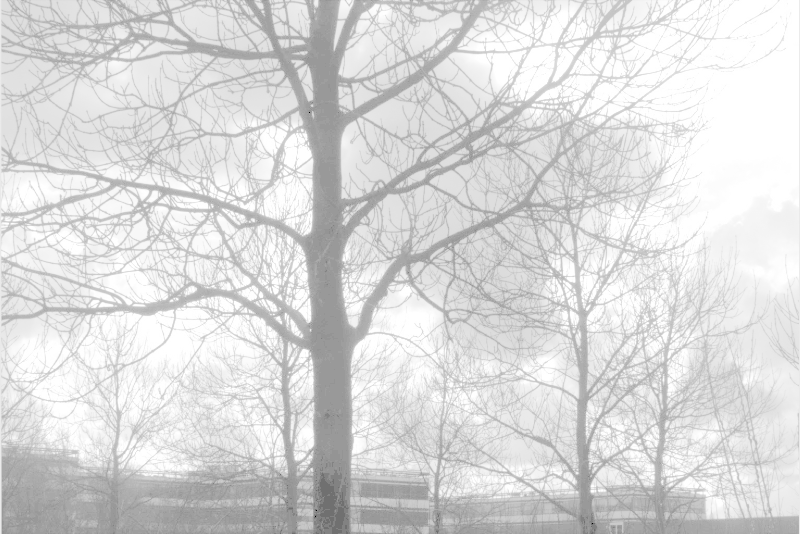}
	\end{minipage}	
	\begin{minipage}{0.01\textwidth}
		
	\end{minipage}
\end{minipage}
\begin{minipage}{\textwidth}
	\begin{minipage}{0.12\textwidth}
		\begin{center}
			PhotoR\\
			Response
		\end{center}
	\end{minipage}	
	\begin{minipage}{0.26\textwidth}
			\includegraphics[width=0.98\textwidth]{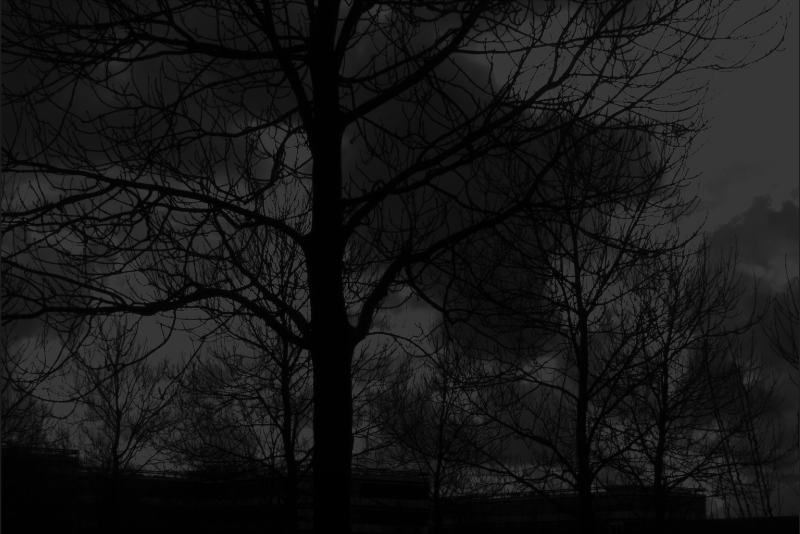}
	\end{minipage}	
	\begin{minipage}{0.26\textwidth}
			\includegraphics[width=0.98\textwidth]{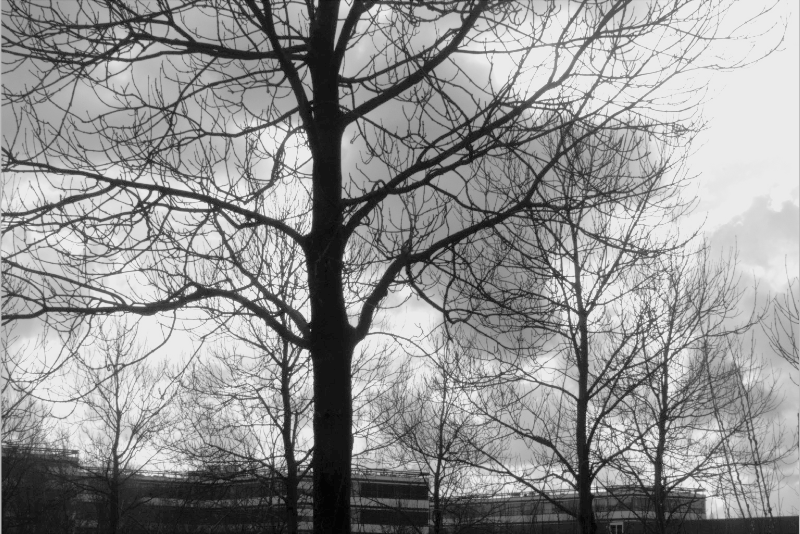}
	\end{minipage}	
	\begin{minipage}{0.26\textwidth}
			\includegraphics[width=0.98\textwidth]{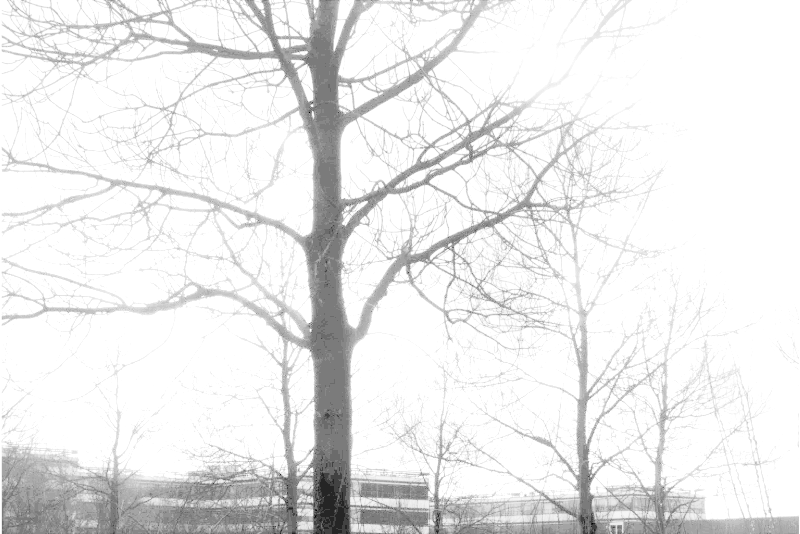}
	\end{minipage}	
	\begin{minipage}{0.01\textwidth}
		
	\end{minipage}
\end{minipage}
\begin{minipage}{\textwidth}
	\begin{minipage}{0.12\textwidth}
		\begin{center}
			PhotoR + Am\\
			Response
		\end{center}
	\end{minipage}	
	\begin{minipage}{0.26\textwidth}
			\includegraphics[width=0.98\textwidth]{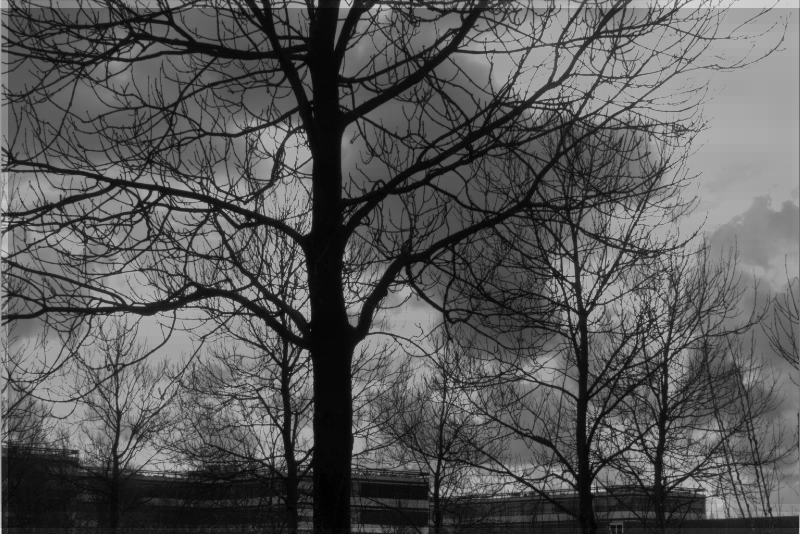}
			\\
			Low Luminance
	\end{minipage}	
	\begin{minipage}{0.26\textwidth}
			\includegraphics[width=0.98\textwidth]{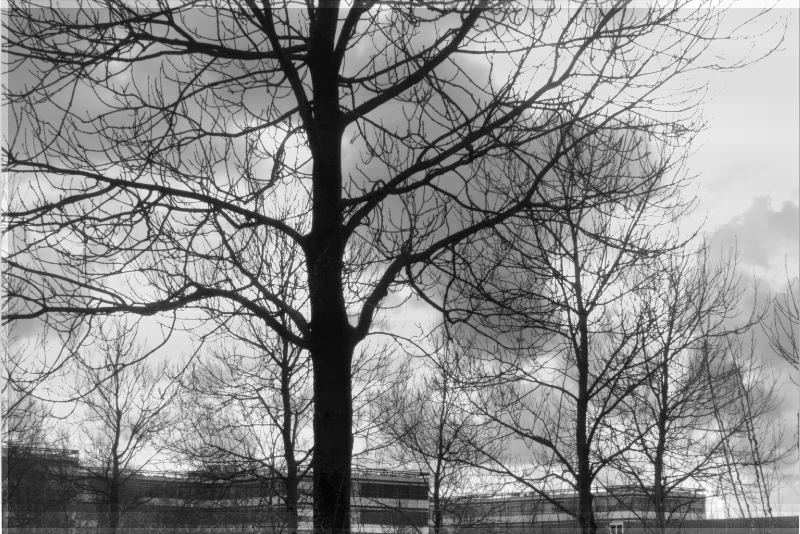}
			\\
			Medium Luminance
	\end{minipage}	
	\begin{minipage}{0.26\textwidth}
			\includegraphics[width=0.98\textwidth]{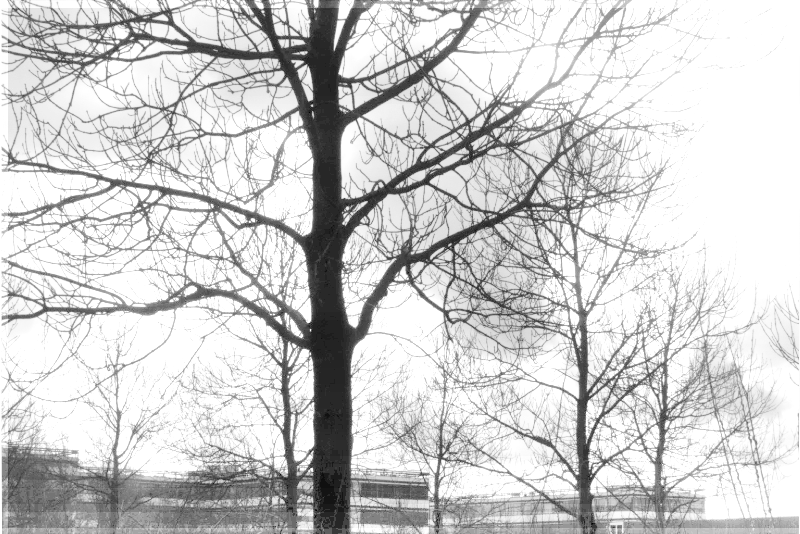}
			\\
			High Luminance
	\end{minipage}	
	\begin{minipage}{0.01\textwidth}
		
	\end{minipage}
\end{minipage}

\caption{Steady state I/O visualization of the spatio-temporal DNP for natural images.
Top Row - Input images. Second Row - Input images visualized on a logarithmic scale. Third Row - Responses  without MVP block. Fourth Row - Responses with  MVP block.
Left Column - Stimuli presented at low luminance, Middle Column - medium luminance, Right Column - high luminance.}
\label{fig:lateral_natural_io}
\end{sidewaysfigure}

We tested the DNP using low, medium and high luminance ($1\times$, $10\times$, and $100\times$) of the original image measured in the number of photons and represented by arbitrary units. They are shown on the top row of Figure~\ref{fig:lateral_natural_io}. The logarithm of the inputs are shown
on the second row. On the third row, we show the response of the DNP 
without the MVP block, \textit{i.e.}, each photoreceptor processes independently a single pixel.
The responses of the full DNP model are shown on the bottom row.

As can be seen from the Figure~\ref{fig:lateral_natural_io}, the
response of the 
DNP with MVP is robust across the different levels of luminance, showing
the best quality for all three luminance levels. By contrast,
the response of the DNP without MVP is either too dark for low contrast
or saturated at high luminance. The contrast within each response is also
significantly worse. It is also not ideal to use only a logarithmic nonlinearity to process the images (second row), as the contrast of the images is not sharp enough.
\end{example}

To conclude we note that, with a simple choice of filters, the spatio-temporal DNP proposed in section~\ref{sec:STDNP} operates in a range of luminance values spanning several orders of magnitude, much like the photoreceptor - amacrine
cell layer. The contrast of the output of the
DNP is not compromised while its range
remains strongly bounded.

\section{Proof for Lemma~\ref{thm:div_iden}}
\label{sec:app_th1}
For the $m^{th}$ trial and the $k^{th}$ time sample, multiplying out the denominator in \eqref{eq:io_n1},  we get, 
\begin{align}
	v^{nm}(t_k) &\left( 1 + \int_\mathbb{D}~h_{1}^{2}(s) u^{nm} (t_k-s) ds + \int_{\mathbb{D}^2}~h_{2}^{2}(s_1,s_2) u_2^{nm}(t_k-s_1, t_k-s_2) \;\;ds_1 ds_2 \right) +\nonumber \\
	&+ v^{nm}(t_k)\left( \int_\mathbb{D}~h_{1}^{3}(s) (\mathcal{P}_1^o v^{nm}) (t_k-s) ds + \int_{\mathbb{D}^2}~h_{2}^{3} (s_1,s_2) (\mathcal{P}_2^o v_2^{nm}) (t_k-s_1, t_k-s_2) \;\;ds_1 ds_2 \right) \nonumber \\
	&= b_1 + \int_\mathbb{D}~h_{1}^{1}(s) u^{nm}(t_k-s) ds + \int_{\mathbb{D}^2}~h_{2}^{1}(s_1,s_2) u_2^{nm}(t_k-s_1, t_k-s_2) \;\;ds_1 ds_2 ,
	\label{eq:sampling}
\end{align}
where $u_2^{nm} (t_1, t_2) \in \mathcal{H}_2$ and $u_2^{nm} (t_1, t_2) = u_1^{nm} (t_1)u_1^{nm} (t_2)$.

With the notation in \eqref{eq:paxes} and after rearranging terms in equation
\eqref{eq:sampling} above, we obtain 
\begin{align}
	v^{nm}(t_k) = 
	 b^1 &+ \int_\mathbb{D}~h_{1}^{1}(s) \phi_{11}^{nmk} ds + \int_\mathbb{D}~h_{1}^{2}(s) \phi_{12}^{nmk} ds 
	 + \int_{\mathbb{D}} ~h_{1}^{3}(s) \phi_{13}^{nmk} ds ~+
	 \label{eq:samp2}\\
	 &+\int_{\mathbb{D}^2}~h_{2}^{1}(s_1,s_2) \phi_{21}^{nmk} \;\;ds_1 ds_2 
	 + \int_{\mathbb{D}^2}~h_{2}^{2}(s_1,s_2) \phi_{22}^{nmk} ds_1 ds_2 
	 + \int_{\mathbb{D}^2}~h_{2}^{3}(s_1,s_2) \phi_{23}^{nmk} ds_1 ds_2 .
	 \nonumber 
\end{align}
With $q^{nmk} = v^{nm}(t_k)$, equation \eqref{eq:samp2} becomes
\begin{equation}
b^1 + \left\langle h_{1}^{1}, \phi_{11}^{nmk} \right\rangle_{\hilbert_1} \!\!+
	 \left\langle h_{1}^{2} , \phi_{12}^{nmk} \right\rangle_{\hilbert_1} \!\!+ 
	 \left\langle h_{1}^{3} , \phi_{13}^{nmk} \right\rangle_{\hilbert_1^o} \!\!
	 \nonumber
	 %\\ &
	 +
	 \left\langle h_{2}^{1}, \phi_{21}^{nmk} \right\rangle_{\hilbert_2} \!\!+ 
	 \left\langle h_{2}^{2}, \phi_{22}^{nmk}  \right\rangle_{\hilbert_2} \!\!\!+
	 \left\langle h_{2}^{3}, \phi_{23}^{nmk}  \right\rangle_{\hilbert_2^o} \!\!\!
	 = q^{nmk} ,
%	\label{eq:norm_inner_prod_form}
\end{equation}
for all $m=1,2,...,M$ and $k=1,2,...,T$.
\qed

\section{Proof of Lemma~\ref{lem:matrix_form}}
\label{sec:lem1}

Let 
\begin{equation}
h^{1}_1 (t) = \sum_{l = -L}^{L} h^1_{1l} \cdot e_{l}(t),
~~~~~
h^{1}_2 (t) = \sum_{l_1 = -L}^{L}\sum_{l_2 = -L}^{L} h^1_{2 l_1 l_2} \cdot e_{l_1}(t_1) \cdot e_{l_2}(t_2),
\label{eq:proj_h11}
\end{equation}
\begin{equation}
h^{2}_1 (t) = \sum_{l = -L}^{L} h^2_{1l} \cdot e_{l}(t),
~~~~~
h^{2}_2 (t) = \sum_{l_1 = -L}^{L}\sum_{l_2 = -L}^{L} h^2_{2 l_1 l_2} \cdot e_{l_1}(t_1) \cdot e_{l_2}(t_2),
\end{equation}
\begin{equation}
h^{3}_1 (t) = \sum_{l = -L^o}^{L^o} h^3_{1l} \cdot e^o_{l}(t),
~~~~~
h^{3}_2 (t) = \sum_{l_1 = -L^o}^{L^o}\sum_{l_2 = -L}^{L} h^3_{2 l_1 l_2} \cdot e^o_{l_1}(t_1) \cdot e^o_{l_2}(t_2),
\end{equation}
and
the $(2L+1)\times1$ vectors
\begin{equation}
\mathbf{h}^1_1 = \begin{bmatrix}
h^1_{1,-L} \\
h^1_{1,-L+1} \\
\vdots \\
h^1_{1,L}  
\end{bmatrix}, \quad\quad
\mathbf{h}^2_1 = \begin{bmatrix}
h^2_{1,-L} \\
h^2_{1,-L+1} \\
\vdots \\
h^2_{1,L}  
\end{bmatrix},
\label{eq:h1_coeff}
\end{equation}
the $(2L^o+1)\times1$ vector
\begin{equation}
\mathbf{h}^3_1 = \begin{bmatrix}
h^3_{1,-L^o} \\
h^3_{1,-L^o+1} \\
\vdots \\
h^3_{1,L^o}  
\end{bmatrix},
\label{eq:h3_coeff}
\end{equation}
the $(2L+1)\times(2L+1)$ Hermitian matrices
\begin{equation}
\mathbf{H}^1_2 = \begin{bmatrix}
h^1_{2,-L,L} & \cdots & h^1_{2,L,L}  \\
\vdots & \ddots & \vdots \\
h^1_{2,-L,-L} & \cdots & h^1_{2,L,-L} 
\end{bmatrix},
\label{eq:h12_coeff}
\end{equation}
\begin{equation}
\mathbf{H}^2_2 = \begin{bmatrix}
h^2_{2,-L,L} & \cdots & h^2_{2,L,L}  \\
\vdots & \ddots & \vdots \\
h^2_{2,-L,-L} & \cdots & h^2_{2,L,-L} 
\end{bmatrix},
\label{eq:h22_coeff}
\end{equation}
and the $(2L^o+1)\times(2L^o+1)$ Hermitian matrix
\begin{equation}
\mathbf{H}^3_2 = \begin{bmatrix}
h^3_{2,-L^o,L^o} & \cdots & h^3_{2,L^o,L^o}  \\
\vdots & \ddots & \vdots \\
h^3_{2,-L^o,-L^o} & \cdots & h^3_{2,L^o,-L^o} 
\end{bmatrix}.
\label{eq:h32_coeff}
\end{equation}

Further, let
\begin{align}
	\mathbf{u}^{nmk} &= \sqrt{S}\begin{bmatrix}
							a^{nm}_{-L} \cdot e_{-L}(t_k)\\
							a^{nm}_{-L+1} \cdot e_{-L+1}(t_k)\\        
							\vdots \\
							a^{nm}_{L} \cdot e_{L}(t_k)
					   \end{bmatrix}, 
					   \qquad \qquad \mathbf{U}_\mathbf{2}^{nmk} = \mathbf{u}^{nmk} \left(\mathbf{u}^{nmk}\right)^\herm ,	
\end{align}
where $a^{nm}_{l}$ 
represents the coefficient of $u_1^{nm}(t)$ w.r.t basis element $e_{l}$ in $\hilbert_1$ such that $u_1^{nm} = \sum_{l=-L}^{L} a^{nm}_{l} e_{l}$, and $(\cdot)^\herm$ represents conjugate transpose,
and let
% $\mathbf{v}^m: \mathbb{D} \rightarrow \mathbb{C}^{2L^o_t+1}$, and $\mathbf{V}_2^m(t):  \mathbb{D} \rightarrow \mathbb{C}^{(2L^o_t+1) \times (2L^o_t+1)}$, be as follows:
\begin{align}
	\mathbf{v}^{nmk} &=  \sqrt{S^o}\begin{bmatrix}
							d^{nm}_{-L^o} \cdot e^o_{-L^o}(t_k)\\
							d^{nm}_{-L^o+1} \cdot e^o_{-L^o+1}(t_k)\\        
							\vdots \\
							d^{nm}_{L^o} \cdot e^o_{L^o}(t_k)
					   \end{bmatrix}, 
					   \qquad \qquad \mathbf{V}_\mathbf{2}^{nmk} = \mathbf{v}^{nmk} \left(\mathbf{v}^{nmk}\right)^\herm ,	
\end{align}
where $d^{nm}_{l}$ represents the coefficient of
$(\mathcal{P}_1^o v^{nm})$ w.r.t basis element $e^o_{l}$ in $\hilbert^o_1$ such that $(\mathcal{P}_1^o v^{nm}) = \sum_{l=-L^o}^{L^o} d^{nm}_{l} e^o_{l}$.

Finally, we define the $(4L+2L^o+4)\times1$ vector
\begin{align}
\boldsymbol\Phi^{nmk} &= \begin{bmatrix}
1 \\ \mathbf{u}^{nmk} \\ -q^{nmk}\mathbf{u}^{nmk} \\ -q^{nmk} \mathbf{v}^{nmk}
\end{bmatrix},
\end{align}
and $(4L+2L^o+3)\times(2L+2L^o+2)$ matrix
\begin{align}
\boldsymbol\Xi^{nmk} &= 
\begin{bmatrix}
\mathbf{U}_2^{nmk} & \mathbf{0}_{(2L+1)\times(2L^o+1)} \\
-q^{nmk}\mathbf{U}^{nmk}_2 & \mathbf{0}_{(2L+1)\times(2L^o+1)} \\
\mathbf{0}_{(2L^o+1)\times(2L+1)} & -q^{nmk} \mathbf{V}_2^{nmk}
\end{bmatrix}.
\end{align}
Finally, we define the $(4L+2L^o+4)\times1$ vector
%$\mathbf{c}_1\in \complex^{2L_t+1} \oplus \complex^{2L_t+1} \oplus \complex^{2L^o_t+1}$ and $\mathbf{C}_2\in  \mathbb{C}^{(2L_t+1) \times (2L_t+1)} \oplus  \mathbb{C}^{(2L_t+1) \times (2L_t+1)} \oplus \mathbb{C}^{(2L^o_t+1) \times (2L^o_t+1)}$ as
\begin{align}
\mathbf{c}_1 = \begin{bmatrix}
 				b^1 &
 				(\mathbf{h}^1_1)^\trans &
 				(\mathbf{h}^2_1)^\trans &
				(\mathbf{h}^3_1)^\trans
 \end{bmatrix}^\trans,
 \end{align}
and the $(4L+2L^o+3)\times(2L+2L^o+2)$ matrix
\begin{align}
\mathbf{C}_2 = \begin{bmatrix}
\mathbf{H}_2^1& \mathbf{0}_{(2L+1)\times(2L^o+1)}\\
\mathbf{H}_2^2 & \mathbf{0}_{(2L+1)\times(2L^o+1)} \\
\mathbf{0}_{(2L^o+1)\times(2L+1)} & \mathbf{H}_2^3
 \end{bmatrix} .
 \label{eq:C2}
 \end{align}

The proof is based on two simple observations. First,
\begin{equation}
\left\langle h_{1}^{1}, \phi_{11}^{nmk} \right\rangle_{\hilbert_1} \!\!= \left\langle
\sum_{l=-L}^{L} h_{1l}^1 \cdot e_l (t) , 
\sum_{l=-L}^{L} a_{l}^{nm} \cdot e_l (t_k - t)
\right\rangle_{\hilbert_1} \!\!\!=
\sum_{l=-L}^{L} h_{1l}^1 a_{l}^{nm} \cdot \sqrt{S}e_l (t_k)
= (\mathbf{h}^1_1)^\trans 
	 \mathbf{u}^{nmk}
\end{equation}
and therefore
\begin{equation}
b^1 + \left\langle h_{1}^{1}, \phi_{11}^{nmk} \right\rangle_{\hilbert_1} \!\!+
	 \left\langle h_{1}^{2} , \phi_{12}^{nmk} \right\rangle_{\hilbert_1} \!\!+ 
	 \left\langle h_{1}^{3} , \phi_{13}^{nmk} \right\rangle_{\hilbert_1^o} \!\! =  \mathbf{c}_1^{\trans} \boldsymbol\Phi^{nmk}.
	 \nonumber
	 \end{equation}
	 Second,
	 \begin{align}
	 \left\langle h_{2}^{1}, \phi_{21}^{nmk} \!\right\rangle_{\hilbert_2} \!\!&=
	 \left\langle \sum_{l_1=-L}^L \sum_{l_2=-L}^L h_{2l_1l_2}^{1} \cdot e_{l_1} (t) \cdot e_{l_2} (s) , \!\!\sum_{l_1=-L}^L a_{l_1}^{nm} e_{l_1} (t_k-t) \!\! \sum_{l_2=-L}^L a_{l_2}^{nm} e_{l_2} (t_k -s) \!\right\rangle_{\hilbert_2}\\
	 &= \sum_{l_1=-L}^L \sum_{l_2=-L}^L h_{2l_1l_2}^{1}
	 a_{l_1}^{nm} a_{l_2}^{nm}
	 \cdot e_{l_1} (t_k) \cdot e_{l_2} (t_k) =
	 \operatorname{Tr}
\left((\mathbf{H}_2^1)^\herm 
\mathbf{U}_2^{nmk}\right)
	 \nonumber
	 \end{align}
	 and, therefore,
	 \begin{equation}
	 \left\langle h_{2}^{1}, \phi_{21}^{nmk} \right\rangle_{\hilbert_2} \!\!+ 
	 \left\langle h_{2}^{2}, \phi_{22}^{nmk}  \right\rangle_{\hilbert_2} \!\!\!+
	 \left\langle h_{2}^{3}, \phi_{23}^{nmk}  \right\rangle_{\hilbert_2^o} =
	 \operatorname{Tr}\left(\mathbf{C}_2^\herm \boldsymbol\Xi^{nmk}\right) .
\end{equation}

\section{Proof of Lemma~\ref{thm:lateral_iden}}
\label{sec:proof2}

For the $m^{th}$ trial, the output of the $n^{th}$ channel,
by multiplying by the denominator of \eqref{eq:stdnp} with $v^{nm}$ sampled at times $t_k$ we obtain 
\begin{align}
        	&v^{nm}(t_k) \Big{(}
        %	\begin{gathered}
				1+\int_\mathbb{D}  h_1^2(s)u^{nm}(t_k-s) ds
        		 + \int_{\mathbb{D}^2}  h_2^2(s_1, s_2)u^{nm}(t_k-s_1) u^{nm}(t_k-s_2) ds_1 ds_2 + 
		 \nonumber \\
        		&+ \int_\mathbb{D}  h_1^3(s)(\mathcal{P}_1^o v^{nm})(t_k-s) ds 
        		+ \int_{\mathbb{D}^2}  h_2^3(s_1, s_2)(\mathcal{P}_1^o v^{nm}) (t_k-s_1) (\mathcal{P}_1^o v^{nm}) (t_k-s_2) ds_1 ds_2 +
		\nonumber \\
        		&+ \sum_{i=1}^N \int_\mathbb{D}  h_1^{i4}(s)(\mathcal{P}_1^o v^{im}) (t_k-s) ds
        		+ \sum_{i=1}^N\sum_{j=1}^N\int_{\mathbb{D}^2}  h_2^{ij4}(s_1, s_2)(\mathcal{P}_1^o v^{im}) (t_k-s_1) (\mathcal{P}_1^o v^{jm})(t_k-s_2) ds_1 ds_2 \Big{)}
		\nonumber \\
		%	\end{gathered}
	&= b^1 + \int_\mathbb{D}  h_1^1(s)u^{nm}(t_k-s) ds + \int_{\mathbb{D}^2}  h_2^1(s_1, s_2) u^{nm}(t_k-s_1) u^{nm}(t_k-s_2) ds_1 ds_2 .
\end{align}
Finally with $v^{nm}(t_k) = q^{nmk}$ and the notation in \eqref{eq:paxes2}, we get
\begin{align}
        	q^{nmk} &= b^1 + \int_\mathbb{D}  h_1^1(s) \phi_{11}^{nmk} (s) ds + \int_{\mathbb{D}^2}  h_2^1(s_1, s_2) \phi_{21}^{nmk} (s_1, s_2) ds_1 ds_2 \nonumber \\ 
	&\qquad \;+\; 
        	      	\int_\mathbb{D}  h_1^2(s) \phi_{12}^{nmk} (s) ds
        		 \;+\; \int_{\mathbb{D}^2}  h_2^2(s_1, s_2) \phi_{22}^{nmk} (s_1,s_2) ds_1 ds_2   \nonumber \\
        		&\qquad+ \int_\mathbb{D}  h_1^3(s) \phi_{13}^{nmk} (s) ds 
        		\;+\; \int_{\mathbb{D}^2}  h_2^3(s_1, s_2) \phi_{23}^{nmk} (s_1,s_2) ds_1 ds_2  \nonumber \\
        		&\qquad+ \sum_{i=1}^N \int_\mathbb{D}  h_1^{i4}(s) \phi_{14}^{inmk} (s) ds 
        		+ \sum_{i=1}^N\sum_{j=1}^N\int_{\mathbb{D}^2}  h_2^{ij4}(s_1, s_2) \phi_{24}^{ijnmk} (s_1,s_2) ds_1 ds_2 ,
		\label{eq:samp3}
\end{align}

\noindent
%Now, let $\zeta^{nm}(s) = v^{nm}(s) - (\mathcal{P}_1^o v^{nm})(s) $. 
and, therefore,
\begin{align}
b^1 &+ 
\left\langle h_{1}^{1}, \phi_{11}^{nmk} \right\rangle_{\hilbert_1} \!\!\!+
	 \left\langle h_{1}^{2} , \phi_{12}^{nmk} \right\rangle_{\hilbert_1} \!\!\!+ 
	 \left\langle h_{1}^{3} , \phi_{13}^{nmk} \right\rangle_{\hilbert_1^o} 
	 + \sum\limits_{i=1}^N \left\langle h_1^{i4} ,
	 \phi_{14}^{inmk}
	 \right\rangle_{\hilbert_1^o} +\\
	 &+
	 \left\langle h_{2}^{1}, \phi_{21}^{nmk} \right\rangle_{\hilbert_2} \!\!\!+ 
	 \left\langle h_{2}^{2}, \phi_{22}^{nmk}  \right\rangle_{\hilbert_2} \!\!\!+
	 \left\langle h_{2}^{3}, \phi_{23}^{nmk} \right\rangle_{\hilbert_2^o}
	 + \sum\limits_{i=1}^N \sum\limits_{j=1}^N \left\langle h_2^{ij4} , 
	 \phi_{24}^{ijnmk}
	 \right\rangle_{\hilbert_2^o} = q^{nm}_k ,
	 \nonumber
\end{align}
for all $m=1,2,...,M$, $k+1,2,...,T$,
and $i,j,n=1,2,...,N$.
\qed

\section{Proof of Lemma~\ref{lem:lem2}}
\label{sec:lem2}

In addition to \eqref{eq:proj_h11}-\eqref{eq:h32_coeff}, let
\begin{equation}
h^{i4}_1 (t) = \sum_{l = -L^o}^{L^o} h^{i4}_{1l} e^o_{l}(t), \quad i =  1,2,\cdots,N,
\end{equation}
\begin{equation}
h^{ij4}_2 (t) = \sum_{l_1 = -L^o}^{L^o}\sum_{l_2 = -L^o}^{L^o} h^{ij4}_{2 l_1 l_2} e^o_{l_1}(t_1)e^o_{l_2}(t_2), \quad i,j = 1,2,\cdots,N,
\end{equation}
and the $(2L^o+1)\times1$ vector
\begin{equation}
\mathbf{h}^{i4}_1 = \begin{bmatrix}
h^{i4}_{1,-L^o} \\
h^{i4}_{1,-L^o+1} \\
\vdots \\
h^{i4}_{1,L^o}  
\end{bmatrix}, \quad i = 1,2,\cdots,N,
\end{equation}
and the $(2L^o+1)\times(2L^o+1)$ matrix
\begin{equation}
\mathbf{H}^{ij4}_2 = \begin{bmatrix}
h^{ij4}_{2,-L^o,L^o} & \cdots & h^{ij4}_{2,L^o,L^o}  \\
\vdots & \ddots & \vdots \\
h^{ij4}_{2,-L^o,-L^o} & \cdots & h^{ij4}_{2,L^o,-L^o} 
\end{bmatrix}, \quad i, j = 1,2,\cdots,N.
\end{equation}
Note that $\mathbf{H}^{ij4}_2, i,j = 1,2,\cdots,N$ are not necessarily Hermitian matrices.
Further, by abuse of notation, let the $(2L+1)\times1$ vector
\begin{equation}
	\mathbf{u}^{nmk} = \sqrt{S}\begin{bmatrix}
							a^{nm}_{-L} \cdot e_{-L}(t_k)\\
							a^{nm}_{-L+1} \cdot e_{-L+1}(t_k)\\        
							\vdots \\
							a^{nm}_{L} \cdot e_{L}(t_k)
					   \end{bmatrix}, 
					   \qquad \qquad \mathbf{U}_\mathbf{2}^{nmk} = \mathbf{u}^{nmk} \left(\mathbf{u}^{nmk}\right)^\herm ,	
\end{equation}
where $a^{nm}_{l}$ represents the coefficient of $u^{nm}(t)$ w.r.t basis element $e_{l}$ in $\hilbert_1$ such that $u^{nm}(t) = \sum_{l=-L}^{L} a^{nm}_{l} e_{l}$, and let the $(2L^o+1)\times1$ vector
\begin{equation}
	\mathbf{v}^{nmk} = \sqrt{S^o}\begin{bmatrix}
							d^{nm}_{-L^o} \cdot e^o_{-L^o}(t_k)\\
							d^{nm}_{-L^o+1} \cdot e^o_{-L^o+1}(t_k)\\        
							\vdots \\
							d^{nm}_{L^o} \cdot e^o_{L^o}(t_k)
					   \end{bmatrix}, 
					   \qquad \qquad \mathbf{V}_\mathbf{2}^{n_1n_2mk} = \mathbf{v}^{n_2mk} \left(\mathbf{v}^{n_1mk}\right)^\herm ,	
\end{equation}
where $d^{nm}_{l}$ represents the coefficient of $(\mathcal{P}_1^o v^{nm})$ w.r.t basis element $e^o_{l}$ in $\hilbert^o_1$ such that $(\mathcal{P}_1^o v^{nm}) = \sum_{l=-L^o}^{L^o} d^{nm}_{l} e^o_{l}$.

Additionally, we define the $\left(1+2(2L+1)+(N+1)(2L^o+1)\right)\times1$ vector
\begin{align}
\boldsymbol\Phi^{nmk} \!\!=\!\! \begin{bmatrix}
1, \!\!&\!\! (\mathbf{u}^{nmk})^\trans\!\!\!, \!\!&\!\!-q^{nmk}(\mathbf{u}^{nmk})^\trans\!\!\!, \!\!&\!\! -q^{nmk} (\mathbf{v}^{nmk})^\trans\!\!\!, \!\!\!&\!\!
 -q^{nmk} (\mathbf{v}^{1mk})^\trans\!\!, \!\!&\!\! -q^{nmk} (\mathbf{v}^{2mk})^\trans\!\!\!, 
 \!\!&\!\! \cdots \!, \!\!&\!\! -q^{nmk} (\mathbf{v}^{Nmk})^\trans
\end{bmatrix}^\trans\!\!&\!\!,
\end{align}
and the $\left(2(2L+1)+(N^2+1)(2L^o+1)\right)\times\left(2L+1+2L^o+1\right)$ matrix
\begin{equation}
\boldsymbol\Xi^{nmk} = 
\begin{bmatrix}
\mathbf{U}_2^{nmk} & \mathbf{0} \\
-q^{nmk}\mathbf{U}^{nmk} & \mathbf{0} \\
\mathbf{0} & -q^{nmk} \mathbf{V}_2^{nnmk} \\
\mathbf{0} & -q^{nmk} \mathbf{V}_2^{11mk} \\
\mathbf{0} & -q^{nmk} \mathbf{V}_2^{12mk} \\
\vdots & \vdots\\
\mathbf{0} & -q^{nmk} \mathbf{V}_2^{NNmk}
\end{bmatrix},
\end{equation}

Finally, we define the $\left(1+2(2L+1)+(N+1)(2L^o+1)\right)\times1$ vector
\begin{align}
\mathbf{c}_1 = \begin{bmatrix}
 				b^1 &
 				(\mathbf{h}^1_1)^\trans &
 				(\mathbf{h}^2_1)^\trans &
				(\mathbf{h}^3_1)^\trans &
				(\mathbf{h}^{14}_1)^\trans &
				(\mathbf{h}^{24}_1)^\trans &
				\cdots &
				(\mathbf{h}^{N4}_1)^\trans 
 \end{bmatrix}^\trans,
 \end{align}
 and the $\left(2(2L+1)+(N^2+1)(2L^o+1)\right)\times\left(2L+1+2L^o+1\right)$ matrix
\begin{align}
\mathbf{C}_2 = \begin{bmatrix}
(\mathbf{H}_2^1)^\trans & (\mathbf{H}_2^2)^\trans 
& \mathbf{0} & \mathbf{0} & \mathbf{0} & \cdots & \mathbf{0} \\
\mathbf{0} & \mathbf{0} & (\mathbf{H}_2^3)^\trans & (\mathbf{H}_2^{114})^\trans  & (\mathbf{H}_2^{124})^\trans  & \cdots & (\mathbf{H}_2^{NN4})^\trans
 \end{bmatrix}^\trans .
 \label{eq:C2_st}
 \end{align}
With the notation above, the proof follows the same steps as the proof of
Lemma~\ref{lem:matrix_form}.

\newpage
\bibliographystyle{unsrt}
\bibliography{reference}

\end{document}